\documentclass[twocolumn]{aastex62}

\usepackage{amsmath}
\usepackage[utf8x]{inputenc}
\usepackage[encapsulated]{CJK}
\usepackage{ucs}
\newcommand{\cntext}[1]{\begin{CJK}{UTF8}{gbsn}#1\end{CJK}}

\newcommand{\angstrom}{\mathrm{\AA}}
\newcommand{\fluxunit}{\rm{erg} \, \rm{cm}^{-2} \, s^{-1} \, {\angstrom}^{-1}}

\newcommand{\ha}{{\rm{H}}\alpha}
\newcommand{\hb}{{\rm{H}}\beta}
\newcommand{\lya}{{\rm{Ly}}\alpha}
\newcommand{\lyb}{{\rm{Ly}}\beta}

\newcommand{\logsha}{\rm log_{10} {[SII}{]}/H\alpha}
\newcommand{\logohb}{\rm log_{10} {[OIII}{]}/H\beta}

\newcommand{\rlya}{{\rm R}_{\lya}}
\newcommand{\ewha}{{\rm EW}_{\ha}}
\newcommand{\ewlya}{{\rm EW}_{\lya}}
\newcommand{\sfrir}{\rm SFR_{IR}}
\newcommand{\sfrha}{{\rm SFR}_{\ha}}
\newcommand{\sfruv}{{\rm SFR}_{\rm UV}}

\newcommand{\frel}{f_{\rm esc,rel}}
\newcommand{\fabs}{f_{\rm esc,abs}}
\newcommand{\ebvmw}{{\rm E(B-V)_{MW}}}
\newcommand{\ebvint}{{\rm E(B-V)_{int}}}

\newcommand{\msun}{{\rm M}_{\odot}}
\newcommand{\zsolar}{{\rm Z}_{\odot}}
\newcommand{\zsubsolar}{{\rm Z}_{1/7\odot}}
\newcommand{\mstar}{{\rm M}_{\star}}

\shorttitle{{[SII]}-Deficiency}
\shortauthors{Wang et al.}

\begin{document}

\title{A New Technique for Finding Galaxies Leaking Lyman-Continuum Radiation: {[SII]}-Deficiency}

\correspondingauthor{Bingjie Wang}
\email{bwang@jhu.edu}

\author[0000-0001-9269-5046]{Bingjie Wang (\cntext{王冰洁}\!)}
\affiliation{Department of Physics \& Astronomy, Johns Hopkins University, Baltimore,
MD 21218, USA}

\author{Timothy M. Heckman}
\affiliation{Department of Physics \& Astronomy, Johns Hopkins University, Baltimore,
MD 21218, USA}

\author{Claus Leitherer}
\affiliation{Space Telescope Science Institute, Baltimore, MD 21218, USA}

\author[0000-0003-2830-0913]{Rachel Alexandroff}
\affiliation{Dunlap Institute for Astronomy \& Astrophysics, University of Toronto,
Toronto, ON M5S 3H4, Canada}

\author{Sanchayeeta Borthakur}
\affiliation{School of Earth \& Space Exploration, Arizona State University, Tempe,
AZ 85287, USA}

\author{Roderik A. Overzier}
\affiliation{Observatorio Nacional, Rio de Janeiro, Brazil}
\affiliation{Institute of Astronomy, Geophysics and Atmospheric Sciences, Department of Astronomy, University of S\~ao Paulo, S\~ao Paulo, SP 05508-090, Brazil}

\begin{abstract}
The source responsible for the reionization of the Universe is believed
to be the population of star-forming galaxies at $z\sim6$ to 12.
The biggest uncertainty concerns the fraction of Lyman-continuum photons
that actually escape from the galaxies. In recent years, several relatively
small samples of ``leaky'' galaxies have been uncovered, and clues
have begun to emerge as to both the indirect signposts of leakiness
and of the conditions/processes that enable the escape of ionizing
radiation. In this paper we present the results of a pilot program
aimed to test a new technique for finding leaky galaxies---using the
weakness of the [SII] nebular emission-lines 
relative to typical star-forming galaxies
as evidence
that the interstellar medium is optically-thin to the Lyman continuum. 
We use the Cosmic Origins Spectrograph on the Hubble Space Telescope to detect significant emerging flux below the Lyman
edge in two out of three [SII]-weak star-forming galaxies at $z\sim0.3$.
We show that these galaxies differ markedly in their properties from
the class of leaky ``Green-Pea'' galaxies at similar redshifts: our
sample galaxies are more massive, more metal-rich, and less extreme
in terms of their stellar population and the ionization state of the
interstellar medium. Like the Green Peas, they have exceptionally high star-formation
rates per unit area. They also share some properties with the known
leaky galaxies at $z\sim3$, but are significantly dustier. Our results
validate a new way to identify local laboratories for exploring the
processes that made it possible for galaxies to reionize the Universe.
\end{abstract}


\section{Introduction}
The Epoch of Reionization (EoR) is the period during which the first
stars are formed and emit light that ionizes the intergalactic medium (IGM).
The history of reionization is primarily inferred from two measurements:
large-scale anisotropies in polarization of the cosmic microwave background (CMB) 
and spectroscopy of distant quasars. 
The CMB is affected by the total column density of free electrons along line of sight.
The parameterization of its Thomson scattering optical depth $\tau$ remains to be the least constrained parameter
in the $\Lambda$CDM model (e.g. \cite{2013ApJS..208...20B, 2018arXiv180706205P}).
Observations of quasar absorption lines via the Gunn-Peterson effect \citep{1965ApJ...142.1633G} 
sets the limit that reionization completes by $z \sim 6$ (e.g. \cite{Fan2006,Fan2006b,Mcquinn2016}, and references therein).

A conventional picture thus depicts the history of reionization as 
early galaxies reionizing hydrogen between $z\sim12$ to 6, and followed by quasars reionizing helium. 
While deep imaging with the Hubble Space Telescope (HST) indicates that
the ultraviolet (UV) luminosity density of early star-forming galaxies
is high enough that they are the best candidates to provide the ionizing
photons necessary for reionizing the Universe (e.g. \cite{Bouwens2015}), 
the fraction of Lyman-continuum (LyC) photons that actually escape from the galaxies 
into the IGM, which is required to be significant (\textgreater\ 0.2), is the biggest uncertainty (e.g. \cite{Robertson2015b}). 
Unfortunately, since the Universe during the EoR is opaque to ionizing photons, direct
observations that access the LyC at these redshifts are impossible. Identifying leaky
galaxies at low redshifts thus becomes an important step in the 
investigation into the physical processes which allow LyC photons to
escape, as well as in identifying indirect observational signposts
of leaky galaxies during the EoR. In addition, we gain sensitivity by looking at local galaxies, which naturally makes 
the relevant analysis easier.

Over the past few years, convincing detections of escaping LyC photons
in a relatively small number of low-redshift starburst galaxies have
emerged \citep{Borthakur2014,Leitherer2016,Izotov2016,Izotov2016b,Izotov2018,Izotov2018a}.
The proposed signposts include a high star formation
rate (SFR) per unit area, strong nebular emission-lines, high flux ratios
of the {[OIII]5007/[OII]}3727 emission lines, and strong $\lya$
emission. In this paper, we present a new and independent signpost
of leakiness that could also be measured by future observations of
galaxies during the EoR by the James Webb Space Telescope (JWST).

The new signpost is the relative weakness of the {[SII]}6717,6731
emission lines, defined with respect to typical star-forming galaxies.
This {[SII]}-deficiency is a tracer of gas that is
optically thin to ionizing radiation, allowing the escape of LyC photons.
Given that the ionization potential for producing SII is only 10.4
eV, which is significantly less than a Rydberg, much of the {[SII]}
emission therefore arises in the warm partially-ionized region near
and just beyond the outer edge of the Stromgren sphere in a classical
HII region. In an HII region that is optically thin to ionizing radiation, this
partially-ionized SII zone is weak or even absent, and the relative
intensity of the {[SII]} emission lines drop significantly as a
result \citep{Pellegrini2012}.

In this paper, we validate this idea using HST far-UV observations
with the Cosmic Origins Spectrograph (COS; \cite{2012ApJ...744...60G}) of a sample of three star-forming
galaxies. The structure of this paper is as follows. In Section~\ref{sec:def}, we begin by detailing our definition of the {[SII]}-deficiency.
In Section~\ref{sec:data}, we summarize the observational data sets, including sample
selection, data processing and analysis, and measured ancillary parameters.
In Section~\ref{sec:results}, we present our results, namely the escape
fractions for the LyC. In Section~\ref{sec:discussion}, we make comparisons
of our galaxies to other known leaky galaxies at both low and high-redshift
selected in other ways, and assess the various indirect indicators
of leakiness. Finally, we summarize our conclusions in Section~\ref{sec:conclusions}.

Throughout we adopt the best-fit cosmological
parameters from the Planck 2018 analysis (their TT,TE,EE+lowE+lensing+BAO case): 
$H_{0}=67.66$ ${\rm km \,s^{-1} \,Mpc^{-1}}$, $\Omega_{M}=0.311$, and $\Omega_{\Lambda}=0.690$ \citep{PlanckCollaboration2018}.

\section{Definition of {[SII]}-deficiency\label{sec:def}}

\begin{deluxetable*}{clccccc}
\tablecaption{Observation logs. \label{tab:sample}}
\tablecolumns{7}
\tablewidth{0pt}
\tablehead{
\colhead{Name} &
\colhead{Galaxy} &
\colhead{$z$} &
\colhead{COS FUV grating} &
\colhead{Exposure time} &
\colhead{COS NUV ACQ image} &
\colhead{Date of HST}\\
\colhead{} & 
\colhead{} & 
\colhead{} &
\colhead{} &
\colhead{(s)} & 
\colhead{exposure time (s)} &
\colhead{observation} 
}
\startdata
J2226 & SDSSJ222634.07-090106.2   & 0.299  & G140L  & 7681.728  & 241  & 2018-05-25\\
J1119 & SDSSJ111905.27+592514.1   & 0.290  & G140L  & 5502.720  & 120  & 2018-09-26\\
J0910 & SDSSJ091021.35+610550.2  & 0.272  & G140L  & 8336.640  & 161  & 2018-09-21\\
J1432 & SDSSJ143256.4+274249.6    & 0.266  & G140L  & 5100.704  & 97  & 2018-06-25\\
J1242 & SDSSJ124206.24+011537.5  & 0.271  & G140L  & 7832.864  & 161  & 2018-08-10\\
\enddata
\end{deluxetable*}

\begin{figure}
  \centering
    \includegraphics[width=0.47\textwidth]{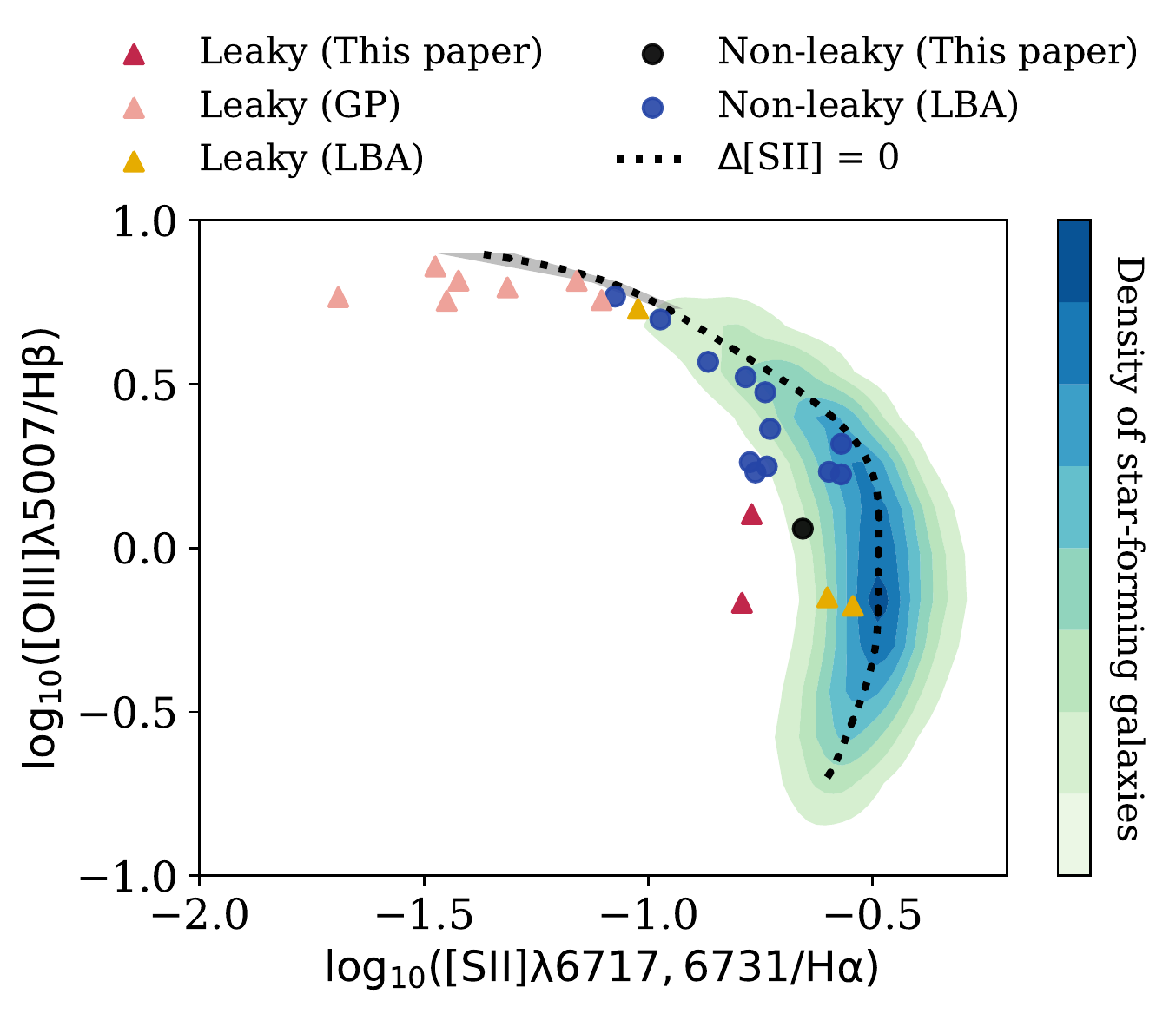}
\caption{This figure is used in defining {[SII]}-deficiency, where the flux
ratio of {[SII]6717,6731}/$\ha$ is plotted against that of {[OIII]5007}/$\hb$.
The contours show the density distribution of the SDSS DR 12 star-forming galaxy sample. 
The black dotted line is fitted to the locus of the peak density of this distribution. The {[SII]}-deficiency is
defined as a galaxy's displacement in log({[SII]}/$\ha$) from this ridge-line.
Uncertainties in the ridge-line are negligible except in the upper left, where they are indicated in grey.
The red triangles represent the two leaky star-forming galaxies of this paper, while the black dot represent the non-leaky one.
Also plotted are leaky Green Pea galaxies in \cite{Izotov2016,Izotov2016b,Izotov2018,Izotov2018a} (pink triangles) and Lyman Break Analogs in \cite{Alexandroff2015} (orange triangles and blue dots), both of which are discussed in Section~\ref{sec:discussion}.
\label{fig:sii}}
\end{figure}

The {[SII]}-deficiency is established with respect to the sample of SDSS DR 12 star-forming galaxies in the plane of {[SII]6717,6731}/$\ha$ {\emph vs.} {[OIII]5007}/$\hb$, as shown in Figure~\ref{fig:sii}. Here we describe the procedure as follows.

First we select all the galaxies classified as ``star forming" in the value added catalog provided by the Portsmouth group \citep{2013MNRAS.431.1383T}, with a signal-to-noise cut of five in the flux measurements.
We then bin the data in $\logohb$ and make a histogram in $\logsha$ for each bin, which is subsequently fitted with a Gaussian (or a skewed Gaussian in a few cases) to determine the peak location. Lastly we perform a polynomial fit to the peaks. This is shown as the black dotted curve in Figure~\ref{fig:sii}. The resulting fitting formula is:
\begin{eqnarray}
y  &=& - 0.487 +  0.014\xi  + 0.028\xi^2 - 0.785\xi^3 \nonumber \\ 
   &  & - 3.870\xi^4 + 0.446\xi^5  + 8.696\xi^6 + 0.302\xi^7 \nonumber \\
   &  & - 6.623\xi^8
\end{eqnarray}
where $\xi$ is the line ratio of $\logohb$, and $y$ is the line ratio of $\logsha$.

We define the {[SII]}-deficiency as a galaxy's displacement in $\logsha$ from the ridge-line, denoted as $\Delta$[SII]. 
Uncertainties in the emission-line ratios for individual galaxies are less than 0.1 dex. Uncertainties in the location of the ridge-line are negligible except where the data are sparse. In these cases, we estimate uncertainties via bootstrap. These are shown in grey in Figure~\ref{fig:sii}.

\section{Data\label{sec:data}}

\subsection{Sample Selection}

In HST program GO-15341 (PI T. Heckman) we observed a sample of five
galaxies selected in the SDSS DR7 plus GALEX GR6 catalogs based on
the following criteria:
\begin{enumerate}
\item A {[SII]}-deficiency relative to normal star-forming galaxies of
at least 0.2 dex as shown in Figure~\ref{fig:sii}. In this paper the value of $\Delta$[SII] for J1242 is just below 0.2 dex. This is because, since the original sample definition, we updated the sample of normal galaxies to SDSS DR 12, which results in a slight change in the ridge-line.
\item A seeing-de-convolved half-light radius of less than $0.5\arcsec$ (typically
smaller than 1 kpc) based on SDSS u-band images. This mimics the small
sizes of galaxies in the EoR.
\item An estimated far-UV flux inside the COS aperture of larger than $2\times10^{-16} \, \fluxunit$
This was derived by using SDSS u-band images to make an aperture correction to the GALEX far-UV flux.
\item Redshifts higher than 0.26. This ensures that the Lyman edge falls
at wavelengths over which COS has high sensitivity (\textgreater
1150 \AA).
\item An SDSS optical spectrum dominated by a starburst (not an active galactic nuclei).
\end{enumerate}
The resulting sample is listed in Table~\ref{tab:sample}. Subsequent
observations with COS show that in the first two galaxies (J2226 and J1119), the far-UV spectrum is dominated by light from a quasar
(a featureless continuum and a strong and very broad $\lya$ emission-line),
even though the SDSS optical spectrum is dominated by a starburst.
We do not discuss these targets further in this paper.
For the three remaining targets, we will demonstrate that they are indeed dominated by starlight in the far-UV by using the fit of Starbust99 (hereafter SB99, \cite{Leitherer1999}) model spectra in Section~\ref{subsec:data-analysis}.

\subsection{Data Processing\label{subsec:data-process}}

All the COS far-UV spectra were obtained using the G140L grating in the
1105 setting. This covers the observed wavelength range from 1110
to 2150 \AA, corresponding to roughly 880 to 1690 \AA\  
in the rest frame. The spectral resolution is about 0.5 \AA.

We first retrieve our COS data from the MAST archive which had been
processed through the standard COS pipeline \verb|CalCOS|. The most
technically challenging part of the data analysis is trying to accurately
subtract the dark counts, which contribute significantly to the net
counts in the region of the LyC. Therefore, following the procedure
in the appendix of \cite{Leitherer2016}, we create a super-dark
image to replace the standard COS pipeline version. A super-dark image
for a given galaxy is obtained by selecting all the COS dark frames
taken within $\pm$ 1 month of the target observing time, and taking an average.
The choice of $\pm$ 1 month is due to the fact that there are temporal fluctuations in the
dark count rate. We therefore turn off the native background correction
in \verb|CalCOS|, and modify the procedure to subtract the super-dark
from the science exposure just before extraction of the spectrum.

By examining the individual dark frames that were used to create a
given super-dark, we estimate that the temporal variations in the
dark count rate leads to an uncertainty in the dark count rate at
the time of the observations of $\pm$ 17\%. This will be one factor
in the accuracy of our measurement of the escaping LyC
flux described below.

\begin{figure*}
\gridline{
\fig{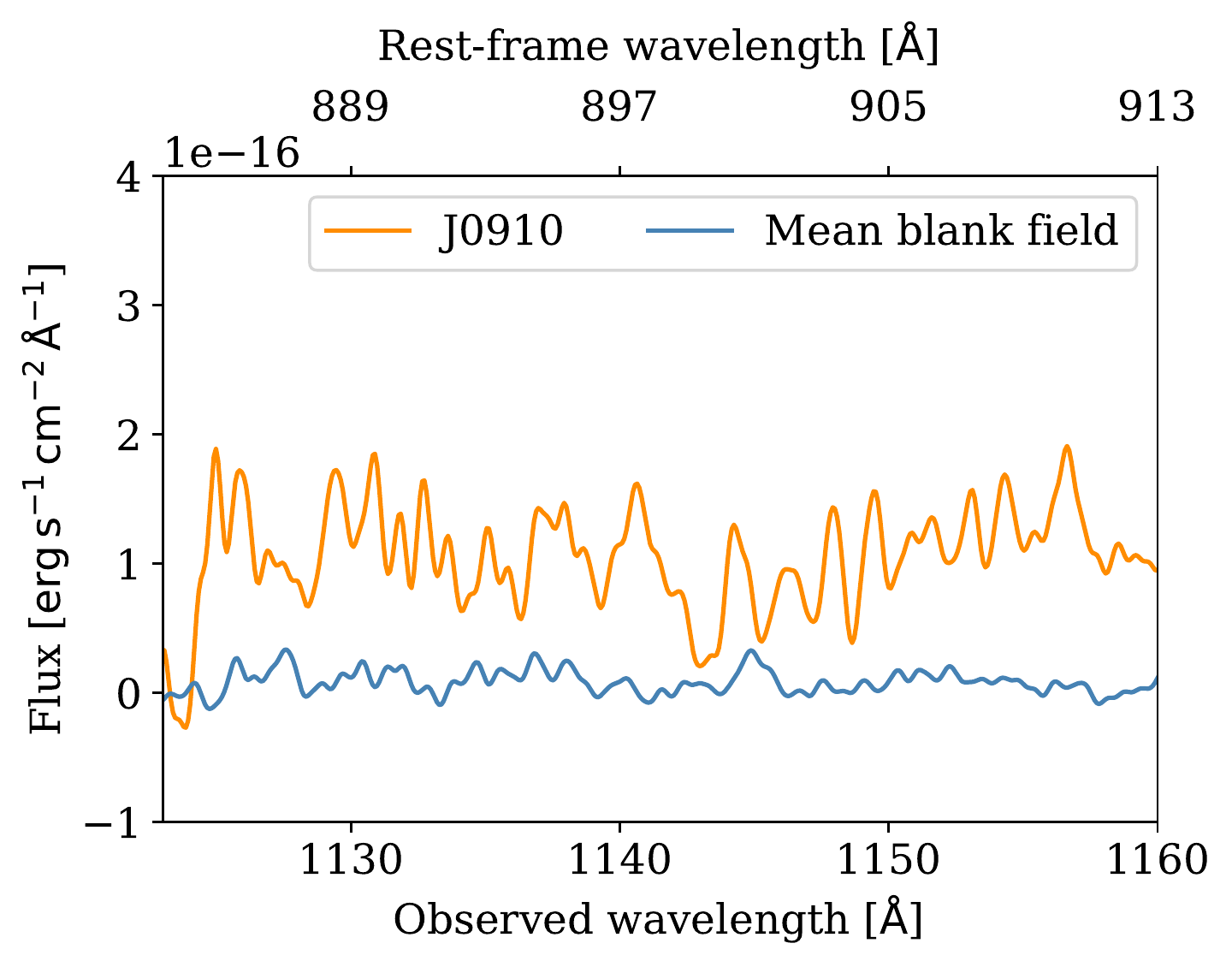}{0.33\textwidth}{(a)}
\fig{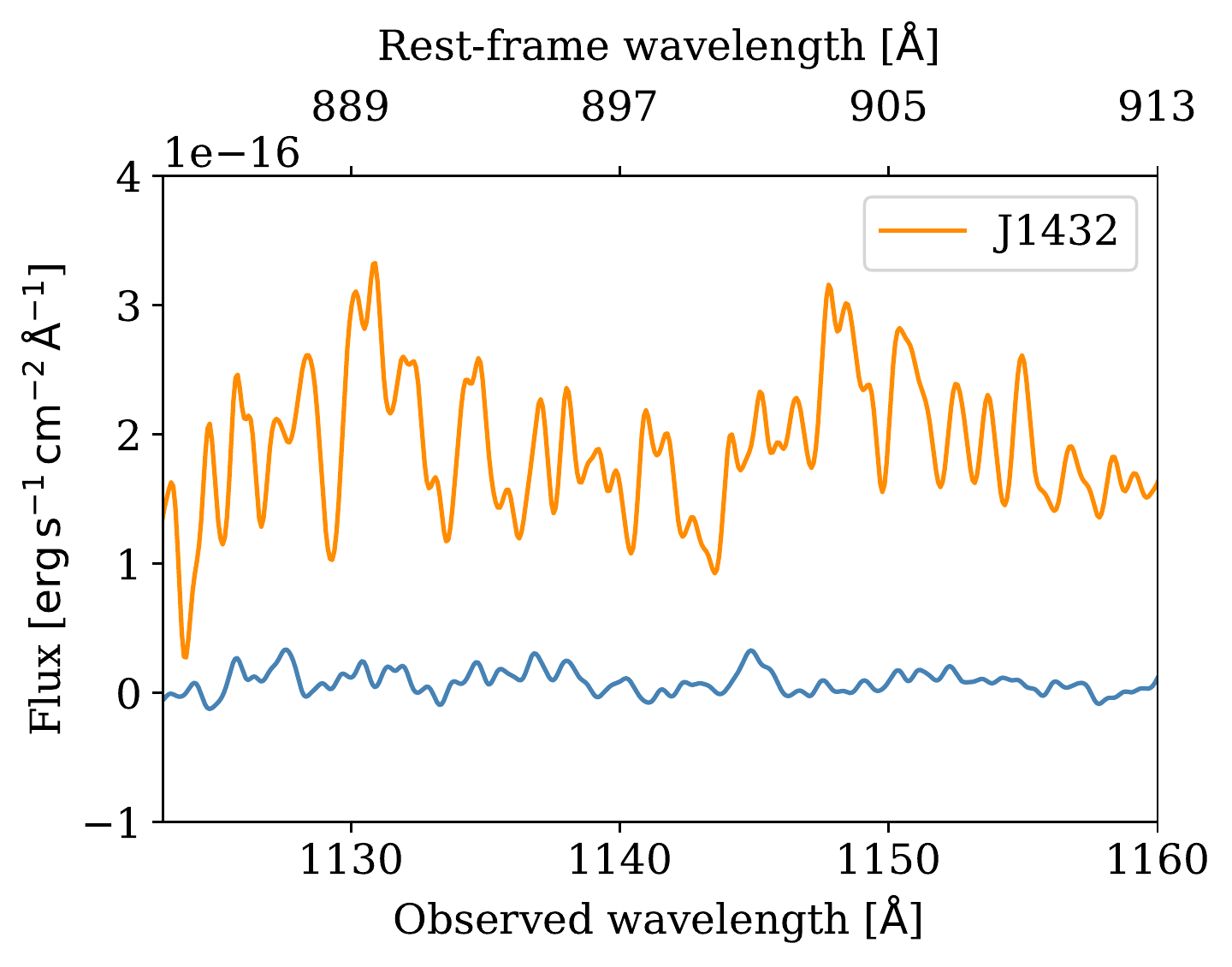}{0.33\textwidth}{(b)}
\fig{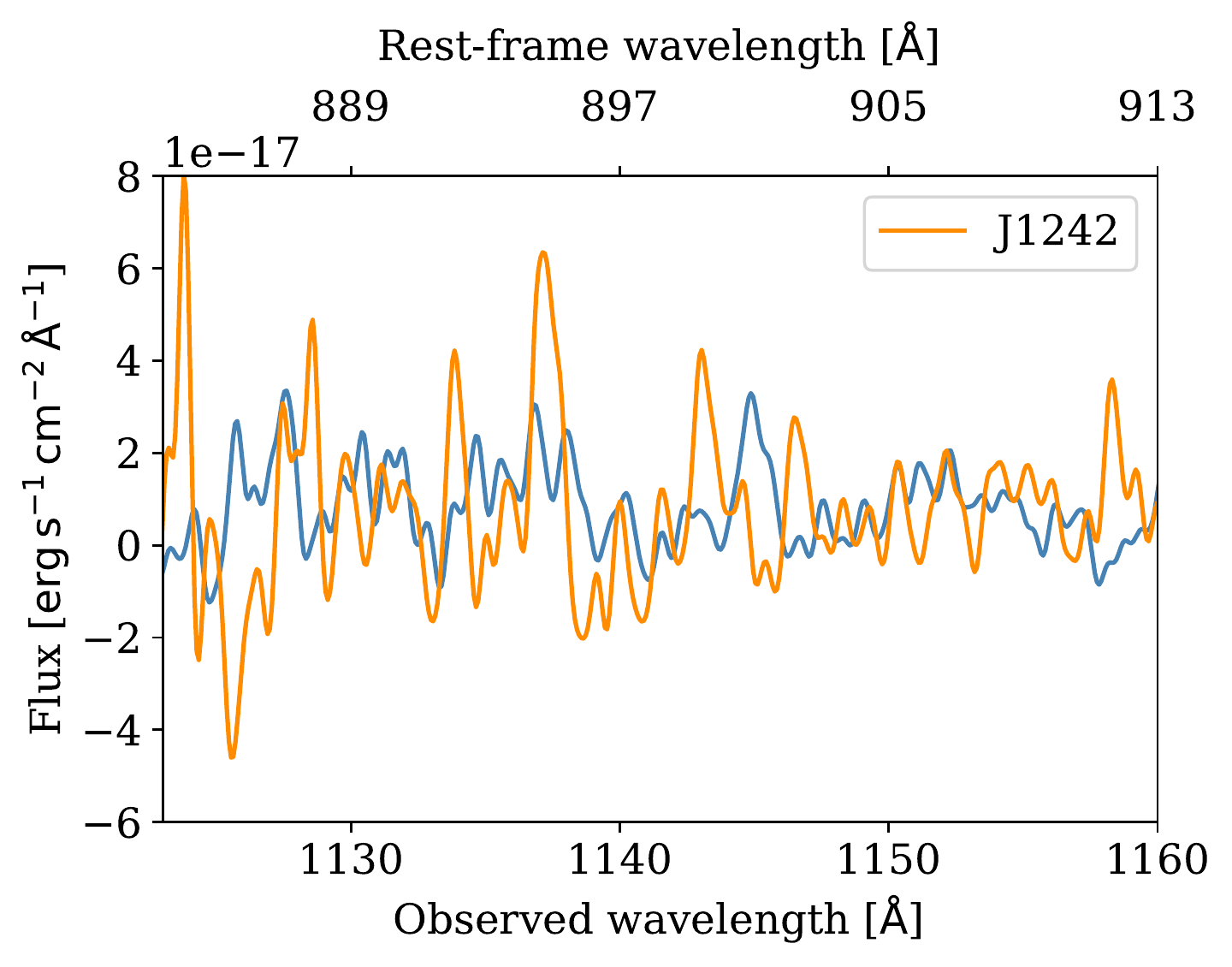}{0.33\textwidth}{(c)}
}
\caption{Observed spectra plotted in the region below the Lyman limit
after super-dark subtraction. The orange lines are the COS spectra
of our three galaxies, while the blue line is the
average of five G140L exposures of blank fields. Note that the blank sky
spectrum shows no contaminating signal.\label{fig:airglow}}
\end{figure*}

We also test possible contamination of the galaxy spectra below
the Lyman edge due to scattered light in the wings of the $\lya$
airglow lines, or other weak emission. To do so, we compare an average
of five G140L exposures of blank fields provided by the COS team with our spectra. This comparison
is shown in Figure~\ref{fig:airglow}, and establishes that there is
no significant sky contamination below the Lyman limit.

\subsection{Data Analysis\label{subsec:data-analysis}}

Given the relatively low signal-to-noise ratio in the extracted spectra,
we smooth all the spectra used with a Gaussian kernel before further analysis. 
The full width at half maximum of the kernel is chosen to be about 0.5 \AA\ 
to reach the native resolution.

For each spectrum, we first correct for Milky Way (MW) extinction in the
observed frame using the reddening law proposed in \cite{Mathis1990},
and $\ebvmw$ taken from the NASA Extragalactic Database for a given position on the sky. We then transform the observed spectra
to the rest frame of the galaxy using SDSS spectroscopic redshifts, conserving the quantity $\lambda F_{\lambda}$.

Synthetic spectra are generated based on stellar evolutionary synthesis
models using SB99.
We produce our models based on a star formation history of a continuous
and constant rate of star formation. The stellar population is parameterized
by a Kroupa initial mass function (IMF) \citep{Kroupa2001}. The stellar population
evolves from the zero-age main sequence using the evolutionary models
of the Geneva Group. The model spectra are described in detail in
\cite{Leitherer2010}. In all, we generate eight sets of SB99
models based on two choices each for burst age ($10^{7}$ and $10^{8}$
years), metallicity (solar or 1/7 solar), and whether or not models
using stellar rotation are employed.

\begin{figure*}
\gridline{
\fig{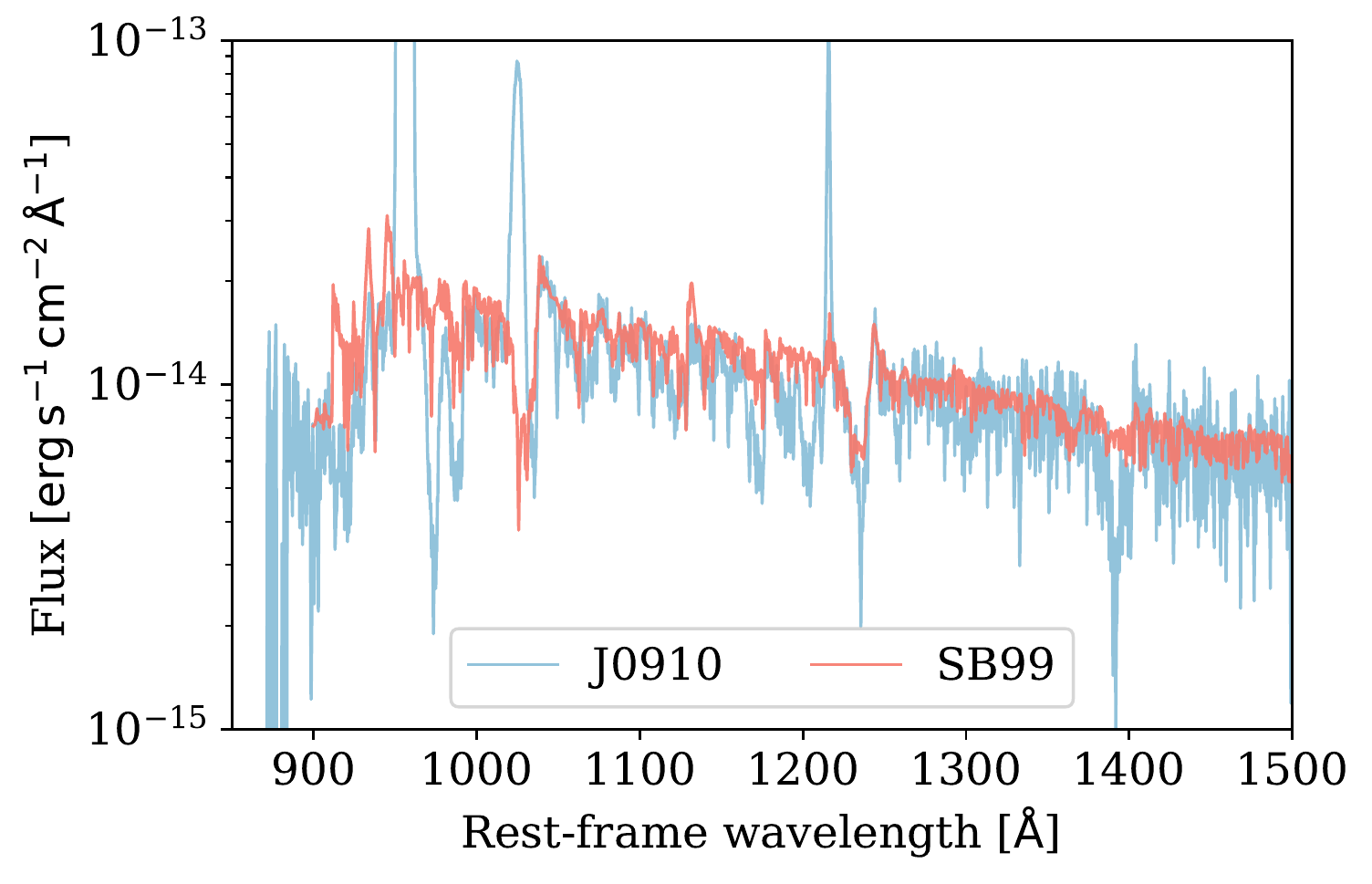}{0.33\textwidth}{(a)}
\fig{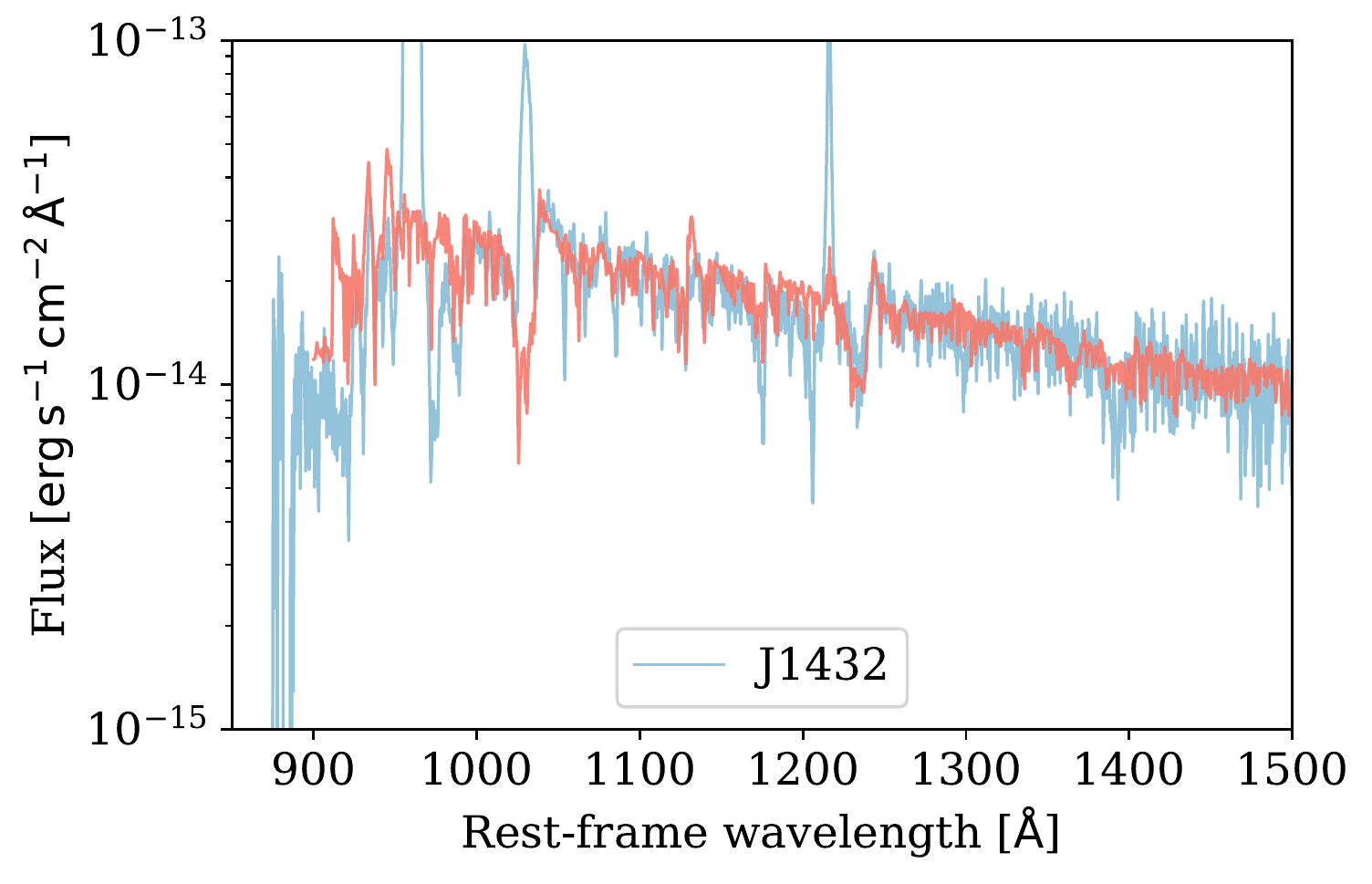}{0.33\textwidth}{(b)}
\fig{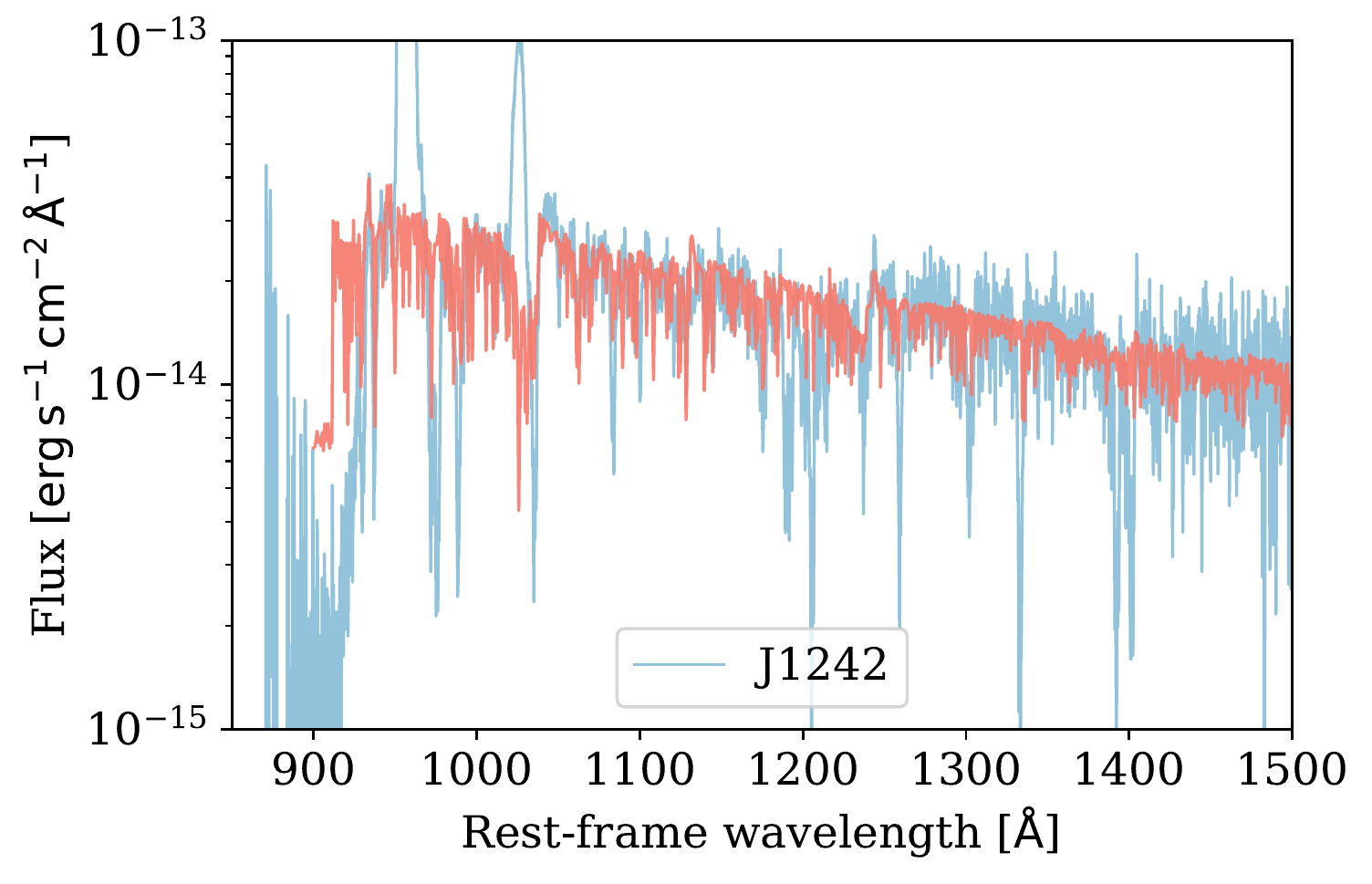}{0.33\textwidth}{(c)}
}
\caption{Spectra of the three star-forming galaxies with Milky Way extinction
and internal extinction removed (in blue), and over-plotted with SB99
best fits (in coral). The extinction values are: (a). $\ebvmw = 0.041$, $\ebvint = 0.239$; (b). $\ebvmw = 0.016$, $\ebvint = 0.243$; and
(c). $\ebvmw = 0.016$, $\ebvint = 0.314$.\label{fig:spectra}}
\end{figure*}

\begin{figure*}
\gridline{
\fig{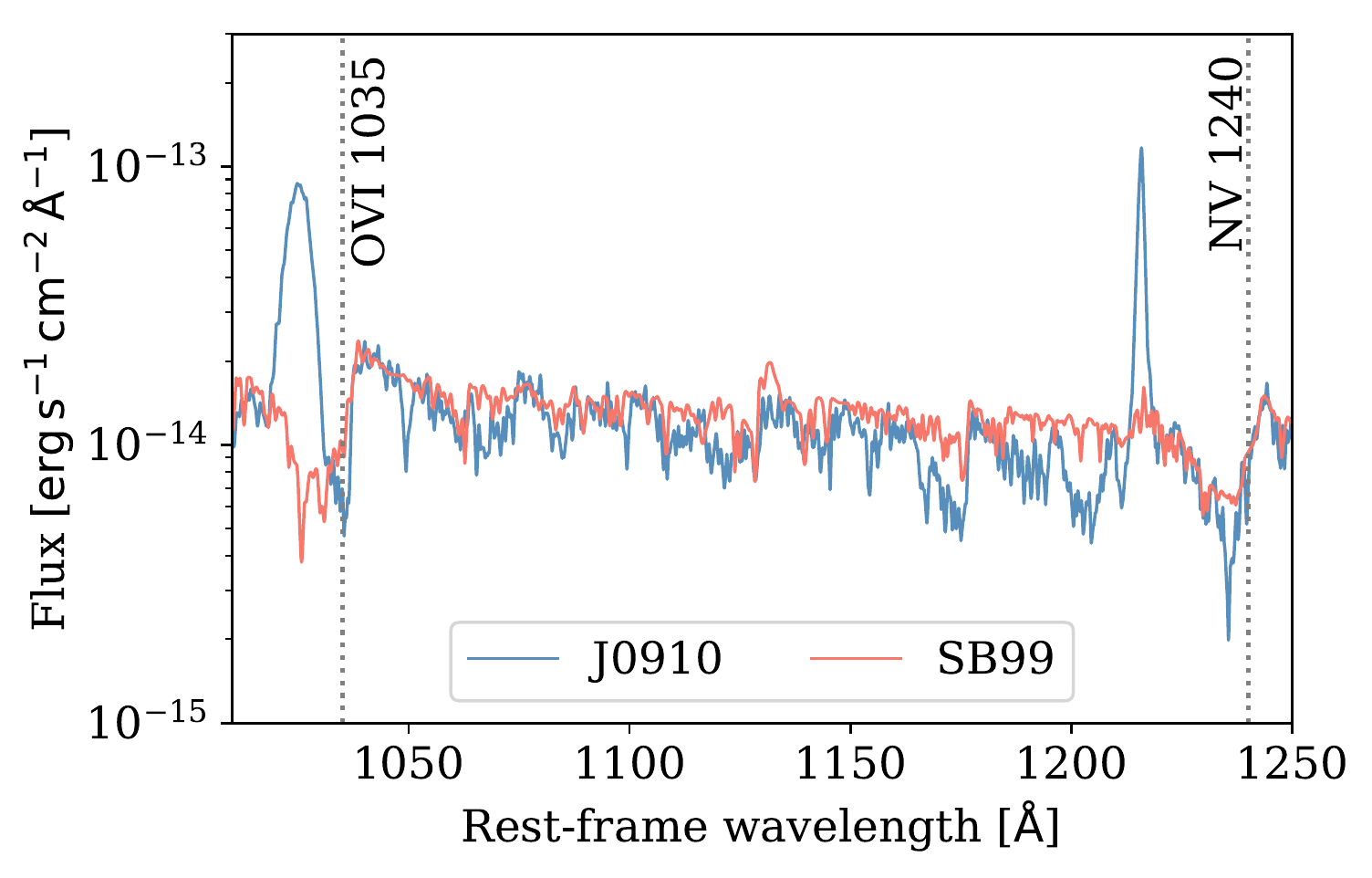}{0.33\textwidth}{(a)}
\fig{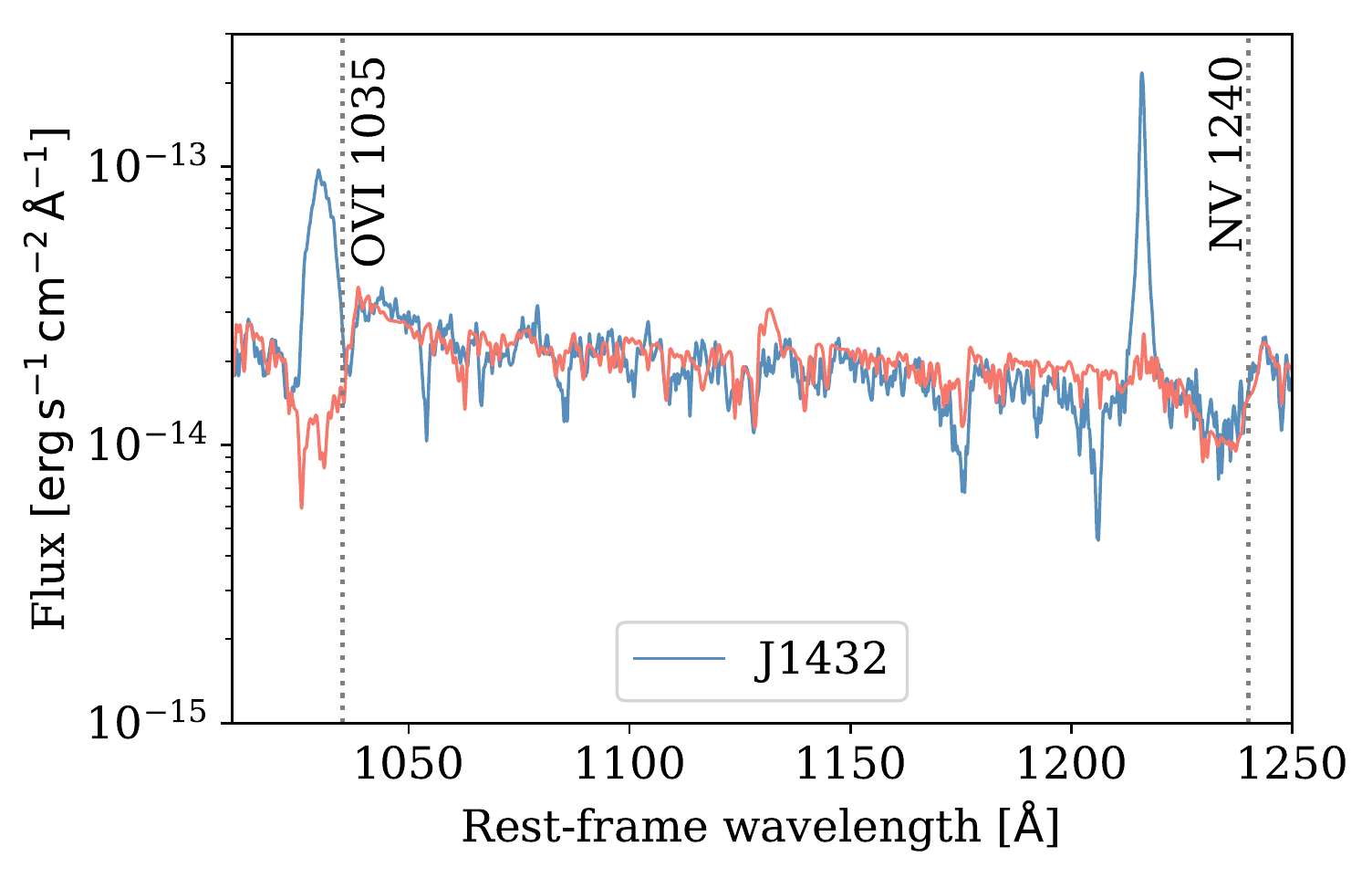}{0.33\textwidth}{(b)}
\fig{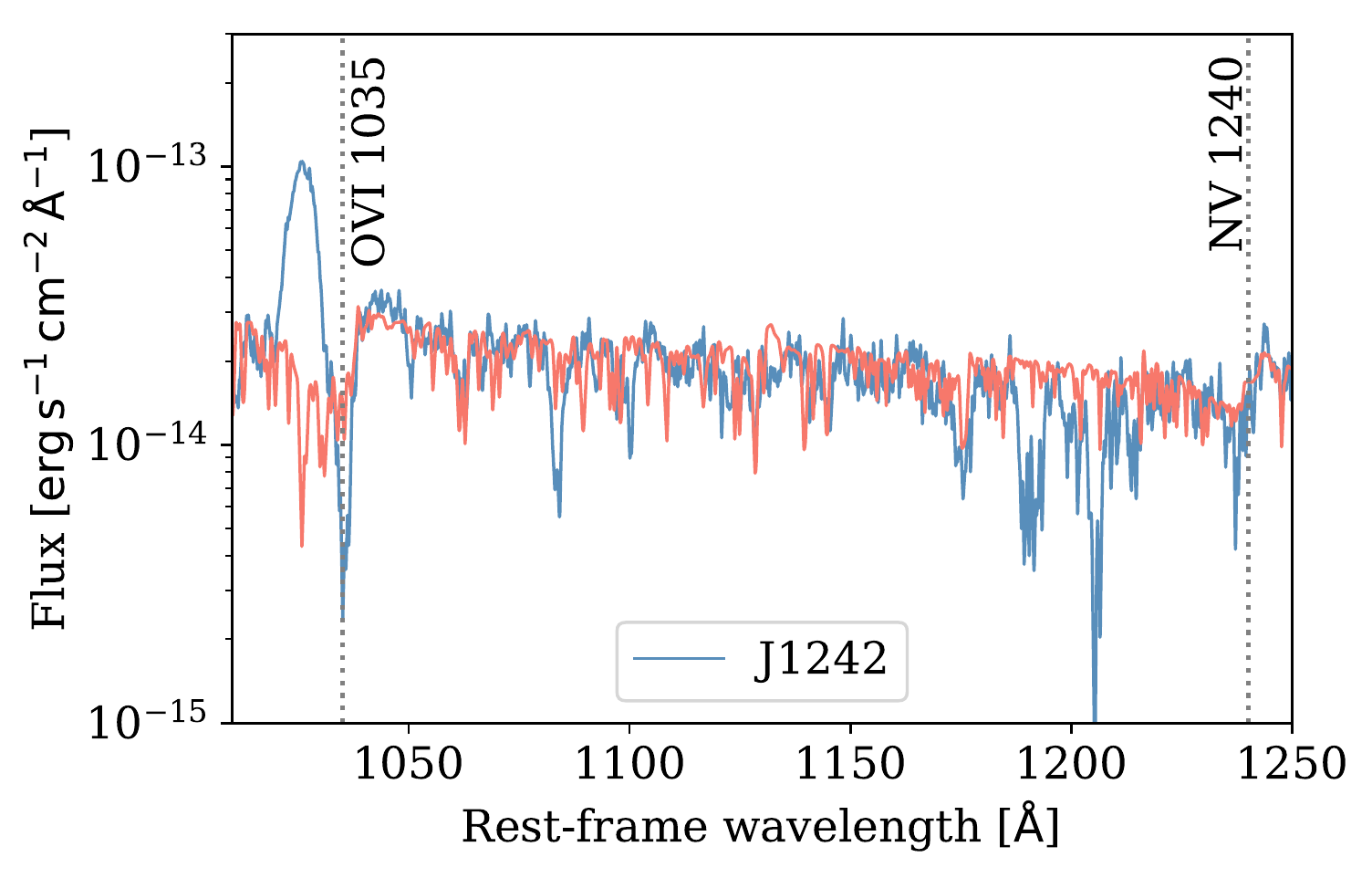}{0.33\textwidth}{(c)}
}
\caption{Same as Figure~\ref{fig:spectra}, but zooming in on the OVI and NV stellar wind lines, which are used for deciding the best-fit SB99 model spectra. The strongest residuals (data minus model) are due to the OI telluric airglow emission, $\lya$ emission, and interstellar absorption-lines. \label{fig:windlines}}
\end{figure*}

A model spectrum is interpolated into the same wavelength array
as its corresponding COS spectrum, and also convolved with the same Gaussian kernel, ensuring that they have the same resolution. 
A best-fit is chosen by eye; more specifically, we closely examine the match between
the synthetic and observed spectra of the two strong stellar wind
features due to OVI 1032,1038 and NV 1238,1242. These P-Cygni features
trace the most massive stars, which are the ones responsible for producing
most of the ionizing continuum. For OVI we could only examine the
redshifted emission component, as the blueshifted absorption is contaminated
by the {[}OI{]} airglow line. From these comparisons, we find that
the best-fits for J0910 and J1432 come from the solar metallicity models that are of $10^{7}$ year ages, and that
incorporate stellar rotation, while J1242 is better fitted with a $10^{8}$-year model. The overall best fits are
shown in Figure~\ref{fig:spectra}, and a zoom-in on these wind lines is shown in Figure~\ref{fig:windlines}.
As seen from the figures, each stellar spectrum alone is a good fit to the data,
and hence we infer that the far-UV light in all three targets is in fact dominated by hot massive stars.
The only stellar feature the model does not fit well is the blend of the CIII 1176 and the CIV/NIV 1169 lines. We are exploring this and will describe the results in a future paper dealing with the stellar populations in these galaxies.

Having chosen a model, we then vary the internal (extragalactic) extinction,
$\ebvint$, as a free parameter until the slope of a given
observed spectrum matches its SB99 model. To do so, we use the extragalactic
reddening law derived in \cite{Calzetti2000}. There is an alternative 
proposed by \cite{Reddy2015,Reddy2016}, which deviates
from the former at short wavelengths ($\lambda<1250$ \AA).
We briefly describe the effect of adopting the Reddy reddening
law in Section~\ref{sec:results} below.

\subsection{Measured Ancillary Parameters\label{subsec:Measured-Ancillary-Parameters}}

\begin{figure*}
\gridline{
\fig{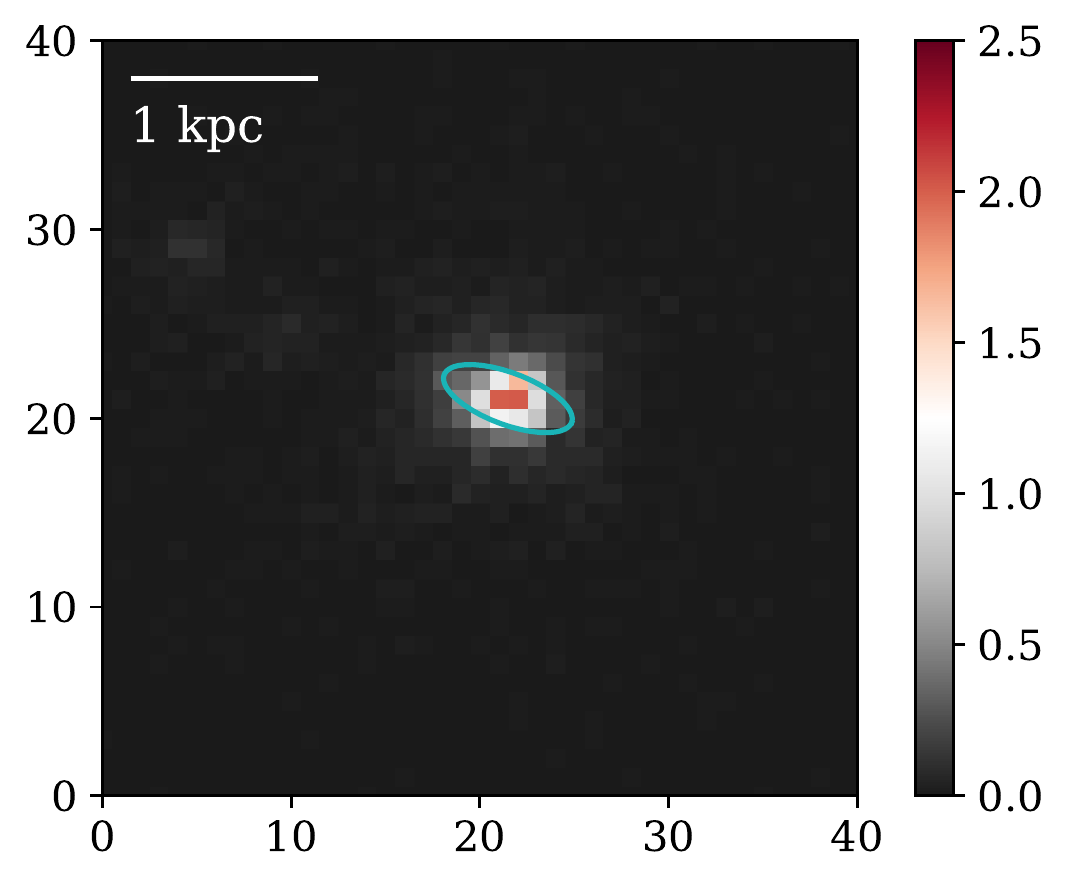}{0.33\textwidth}{(a)}
\fig{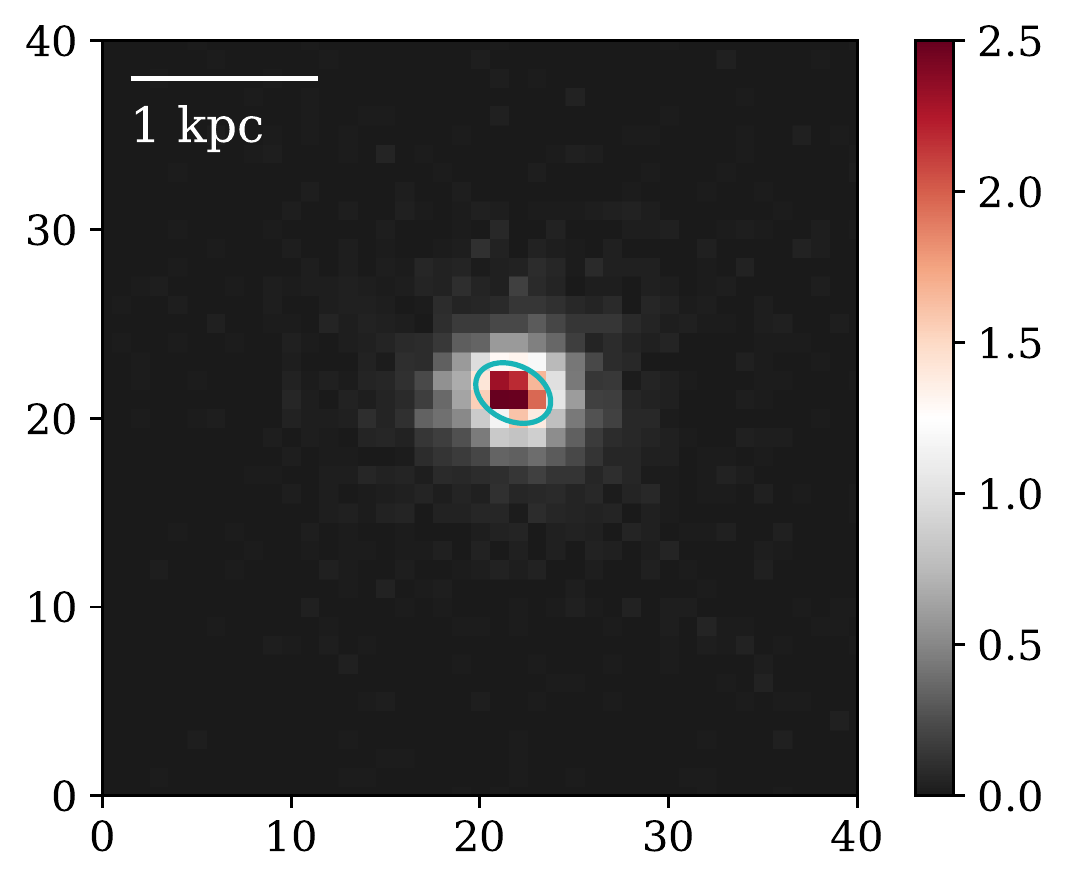}{0.33\textwidth}{(b)}
\fig{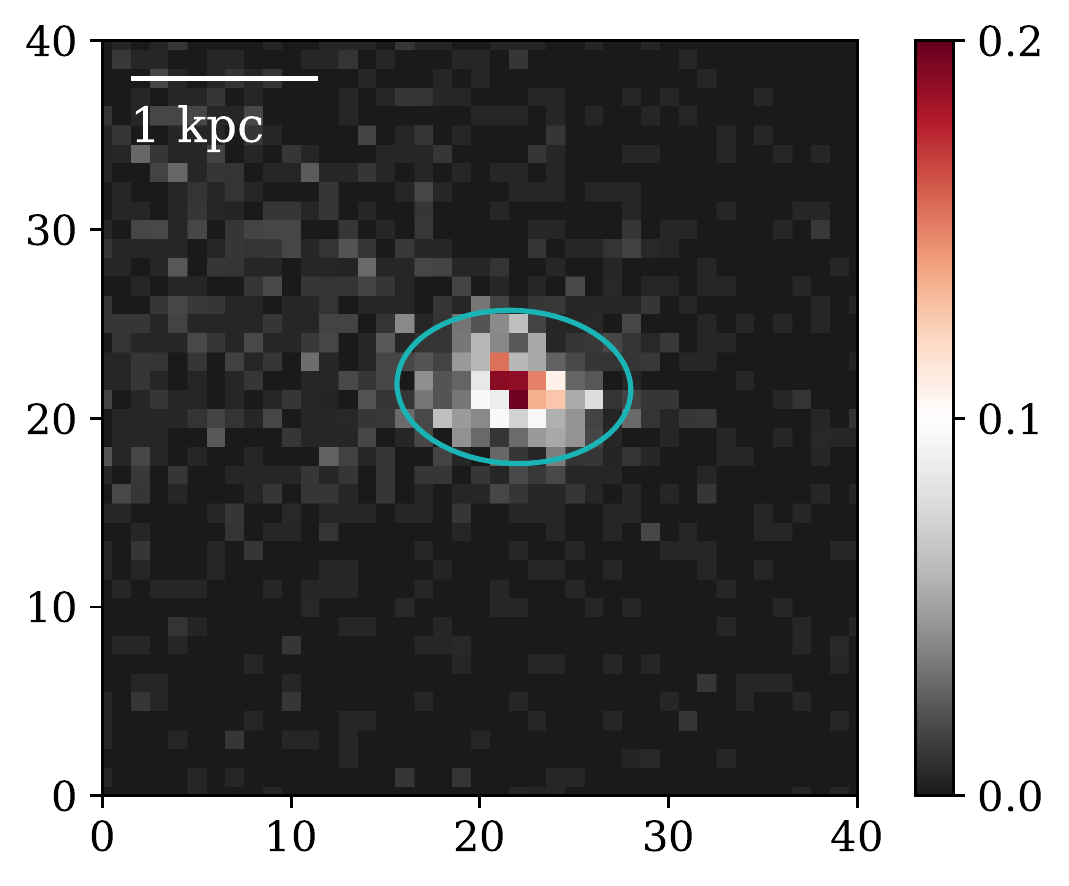}{0.33\textwidth}{(c)}
}
\caption{COS near-UV acquisition images of the three [SII]-weak star-forming galaxies: (a) J0910; (b) J1432; and (c) J1242. Also over-plotted in turquoise are the ellipses which enclose half of the total near-UV emitted light. The images are 1.032\arcsec\ by 1.032\arcsec, and the color bars indicate counts per second.\label{fig:acq}}
\end{figure*}

In this section we list important ancillary parameters, and describe
how they are determined. The values are all listed in Table~\ref{tab:measured}.

\begin{deluxetable*}{cccccccc}
\tablecaption{Measured ancillary parameters.\label{tab:measured}}
\tablecolumns{8}
\tablewidth{0pt}
\tablehead{
\colhead{} &
\colhead{$\sfruv$} &
\colhead{$\sfrha$} &
\colhead{$\sfrir$} &
\colhead{$\rm A (\ha)$} &
\colhead{$r_{50}$} &
\colhead{$\sfrir$/A} &
\colhead{$\rm M_\star$} \\
\colhead{} &
\colhead{$(\msun {\rm {yr}^{-1})}$} &
\colhead{$(\msun {\rm {yr}^{-1})}$} &
\colhead{$(\msun {\rm {yr}^{-1})}$} &
\colhead{} &
\colhead{(kpc)} &
\colhead{$(\msun {\rm \,{yr}^{-1}\,{\rm kpc}^{-2}})$} &
\colhead{($\rm log_{10}\msun$)} 
}
\startdata
J0910  & 128  & 35  & $125\pm11$ & 1.24 & 0.22 & 394 & 10.44  \\
J1432  & 209  & 19  & $134\pm10$ & 0.75 & 0.17  & 705 & 10.54   \\
J1242  & 100  & 21  & $55\pm10$   & 0.96 & 0.50  & 34  & 10.38  \\
\hline
\hline
 & $\ewha$ & $\ewlya$ & $\rlya$ & $\Delta${[SII]} & {[OIII]}/{[OII]} &  $12+{\rm log}_{10}({\rm O/H})$ &  \\
 &  (\AA)   &  (\AA)      &             & (dex)               &                       &  & \\
\hline
J0910 & 138  & 21.84 & 0.75  & -0.30  & 1.29  & 8.66  & \\
J1432 & 113  & 24.55 & 0.44  & -0.28  & 1.57  & 8.60  & \\
J1242 & 125  & N/A    & N/A   & -0.17  & 1.42  & 8.52  & \\
\enddata
\end{deluxetable*}

We measure the star formation rates (SFRs) in three ways. In all cases we use the same IMF as that used in our SB99 fit (see above). $\sfruv$ is inferred from COS UV data by taking a ratio between a dereddened galaxy flux spectrum and a SB99 spectrum generated assuming a SFR of 1 $\msun{\rm yr^{-1}}$. $\sfrir$ is calculated by using the WISE IR data at 12 and 22 $\mu$m \citep{2010AJ....140.1868W} to estimate the rest-frame 24 $\mu$m luminosity, and then using the relation given in \cite{Kennicutt2012}. This has the advantage of being independent of any uncertain correction to the UV fluxes. $\sfrha$ is calculated from extinction-corrected fluxes. The MPA-JHU catalog provides the fluxes of $\ha$ and $\hb$.
We calculate $E(\beta-\alpha)$, defined as $E(\beta-\alpha)=A(\ha)-A(\hb)$
with $A$ being the extinction in magnitude, as:
\begin{subequations}
\begin{eqnarray}
E(\beta-\alpha) & = & 2.5\,{\rm log}_{10}[F(\ha)_{{\rm obs}}/F(\hb)_{{\rm obs}}] \nonumber \\
 & &-2.5\,{\rm log}_{10}[F(\ha)/F(\hb)]
\end{eqnarray}
Assuming a temperature of $10^4$ K, which translates to an intrinsic ratio of $F(\ha)/F(\hb)=2.86$, the extinction
in magnitude for $\ha$ is then

\begin{equation}
A(\ha)=2.29\,E(\beta-\alpha)
\end{equation}
And so finally we have the extinction-corrected $\ha$ flux as:

\begin{equation}
F(\ha)_{{\rm corr}}=10^{0.4A(\ha)}\,F(\ha)_{{\rm obs}}
\end{equation}
\end{subequations}
Following Table 1.1 in \cite{Calzetti2011}, we estimate ${\rm SFR}(\ha)$
in units of $\msun{\rm yr}^{-1}$ via

\begin{equation}
{\rm SFR}(\ha)=5.5\times10^{-42}\,L(\ha)\label{eq:sfr_ha}
\end{equation}
where $L$ is the luminosity in erg ${\rm s}^{-1}$.

After examining the COS near-UV acquisition images as shown in Figure~\ref{fig:acq}, we find that all
targets are well located inside the SDSS and COS apertures, which
are taken to be $1.5\arcsec$ and $1.2\arcsec$ respectively. We therefore do not
apply any aperture corrections to SFRs. We also note that the fluxes in the images are consistent with the GALEX near-UV flux.

Additionally we use the COS near-UV images to compute the half-light radius for a given galaxy by finding an ellipse that encloses half of the total near-UV emitted light; the listed value for $r_{50}$ is therefore $(a_{50} b_{50})^{1/2}$ in kpc. 
During the process, the background is estimated from the mean of an annulus of $\rm r_{in}=0.9\arcsec$ and $\rm r_{out}=1.1\arcsec$. A small correction for the effect of the PSF is also applied.

The values for the rest-frame equivalent
width of the $\ha$ emission line are taken from the MPA-JHU catalog,
and the stellar masses are taken from the median of the corresponding
PDF in the same catalog.

The oxygen abundance of the interstellar medium (ISM) in each galaxy is estimated following \cite{Pettini2004}:
\begin{subequations}
\begin{equation}
12+{\rm log}_{10}({\rm O/H})=8.73-0.32\times{\rm O}3{\rm N}2
\end{equation}
where
\begin{equation}
{\rm O}3{\rm N}2={\rm log}_{10}\frac{[{\rm O_{III}}]\lambda5007/\hb}{[{\rm N_{II}}]\lambda6584/\ha}
\end{equation}
\end{subequations}
This relation is valid for $-1<$O3N2$<1.9$. Since the wavelength
of $\ha$ is close to {[}NII{]} and $\hb$ is close to {[OIII]},
this method is insensitive to dust extinction. Then we use the conversion:
$12+{\rm log}_{10}({\rm O/H})=8.7$ corresponding to solar metallicity.

To characterize the $\lya$ line, we use the following procedure.
Each observed galaxy spectrum is first normalized by fitting a second-order
polynomial function to the continuum and the spectrum is divided by
this function. We do the same for the corresponding best-fit SB99
spectrum, which is then subtracted from the normalized
galaxy spectrum to remove the stellar spectral component. Lastly we add a value
of $1$ to this difference spectrum to produce a normalized spectrum
with stellar features removed.

To measure the $\lya$ equivalent widths, we fit a (multi-component-)
Gaussian. We estimate that the resulting equivalent widths have errors
on the order of 10\%--15\% dominated by systematics in the polynomial
fit to the continuum emission and the subtraction of SB99 models.

Next, we use the starlight-subtracted spectra to quantify the different $\lya$
profile shapes using the parameter $\rlya$, which is defined to be
the ratio of the equivalent width of the blueshifted portion of the
profiles to that of the redshifted portion. We define the equivalent
width for emission to be positive, and for absorption to be negative.
Therefore, a negative $\rlya$ indicates blueshifted absorption and
redshifted emission (i.e. a traditional P-Cygni profile) while a positive
value for $\rlya$ indicates significant blueshifted emission.

\section{Results\label{sec:results}}

\begin{deluxetable*}{ccccccc}
\tablecaption{Measurements of observed flux densities used in quantifying the escape fractions. The LyC ranges are wavelength ranges over which an average is taken in calculating $F_{900}$ and $F_{900^-}$. The first uncertainties in $F_{900}$ are estimated from Poisson statistics, and the second ones are from dark fluctuations. \label{tab:flux}}
\tablecolumns{7}
\tablewidth{0pt}
\tablehead{
\colhead{} &
\colhead{LyC range} &
\colhead{$F_{900}$\tablenotemark{a}} &
\colhead{$F_{900}$\tablenotemark{b}} &
\colhead{$\frac{F_{900}}{F_{1500}}$\tablenotemark{b}} &
\colhead{$\left( \frac{F_{900^-}}{F_{900^+}}\right)_{\rm obs}$\tablenotemark{c}} &
\colhead{$\left( \frac{F_{900^-}}{F_{900^+}}\right)_{\rm obs}$\tablenotemark{d}} \\
\colhead{} &
\colhead{(\AA)} &
\colhead{($\times 10^{-16} \fluxunit$)} &
\colhead{($\times 10^{-16} \fluxunit$)} &
\colhead{} &
\colhead{} &
\colhead{}
}
\startdata
J0910  & 885 -- 910 & $1.38 \pm 0.17 \pm 0.05$ & 2.16 &0.38 &0.538 & 0.482\\
J1432  & 888 -- 910 & $2.52 \pm 0.19 \pm 0.08$ & 2.99 & 0.32 &0.460 & 0.406\\
J1242  & 885 -- 910 & $0.10 \pm 0.16 \pm 0.08$ & 1.04 & 0.02 & 0.046 &0.039
\enddata
\tablenotetext{a}{Uncorrected for extinction.}
\tablenotetext{b}{Corrected for MW extinction only.}
\tablenotetext{c}{Corrected for MW and internal extinctions, assuming extragalactic reddenning law in \cite{Calzetti2000}.}
\tablenotetext{d}{Same as c, but assuming extragalactic reddenning law in \cite{Reddy2015,Reddy2016}.} 
\end{deluxetable*}

\begin{deluxetable*}{ccccccc|c}
\tablecaption{Relative and absolute escape fractions. The measurements quoted for J1242 are upper limits inferred from a 3$\sigma$ limit on dark fluxes. The first uncertainties are estimated from Poisson statistics, and the second ones are from dark fluctuations.\label{tab:fesc}}
\tablecolumns{8}
\tablewidth{0pt}
\tablehead{
\colhead{ } &
\colhead{$\ebvint$\tablenotemark{a}} &
\colhead{$\frel$} &
\colhead{$\fabs$} &
\colhead{$\ebvint$\tablenotemark{b}} &
\colhead{$\frel$} & 
\colhead{$\fabs$} &
\colhead{$\fabs$\tablenotemark{c}} \\
\colhead{} &
\colhead{} &
\colhead{($\times 10^{-2}$)} &
\colhead{($\times 10^{-2}$)} &
\colhead{} &
\colhead{($\times 10^{-2}$)} &
\colhead{($\times 10^{-2}$)} &
\colhead{($\times 10^{-2}$)}
}
\startdata
J0910  & 0.239 & $93.3^{+10.2+3.2}_{-10.0-3.1}$   & $3.3\pm 0.4 \pm0.1$  & 0.257 & $83.6^{+9.1+2.8}_{-8.9-2.8}$  & $4.0\pm0.4\pm0.1$ & $3.5\pm0.5\pm0.3$ \\
J1432  & 0.243 & $79.8^{+5.6+2.3}_{-5.5-2.3}$  & $2.7\pm0.2\pm0.1$  & 0.252 & $70.4^{+6.6+2.1}_{-6.2-2.1}$  & $3.5\pm0.3\pm0.1$ & $4.1\pm0.4\pm0.3$\\
J1242  & 0.314 & \textless 28.3  & \textless 0.4 & 0.325 & \textless 24.4 & \textless 0.5 & \textless 0.7\\
\enddata
\tablenotetext{a}{Assuming reddenning law in \cite{Calzetti2000}.}
\tablenotetext{b}{Assuming reddenning law in \cite{Reddy2015,Reddy2016}.} 
\tablenotetext{c}{Obtained by taking the ratio between MW extinction-corrected $(F_{900})_{\rm obs}$ and $(F_{900})_{\rm int}$ inferred from SB99 given $\sfrir$.}
\end{deluxetable*}

As can be seen in Figure~\ref{fig:airglow}, we detect a significant
flux below the Lyman edge in J0910 and J1432, and measure only
an upper limit in J1242. To characterize this emission,
we take the mean of flux densities, uncorrected for extinction, in a spectral window from $\sim$ 885 to 910 \AA. Resulting values are listed in Table~\ref{tab:flux} as $F_{900}$. The exact spectral windows for each of the galaxies are also listed in Table~\ref{tab:flux} under the column LyC range. These particular choices are motivated by avoiding the detector edge where dark count rates increase significantly.
The errors quoted account for both the statistical (Poisson) errors,
which are extracted from the corresponding \verb|x1d| files, and
the systematic errors associated with dark subtraction.

In the following paragraphs, we consider three ways to measure the escape fraction, each
with advantages and disadvantages. Relevant measurements of flux densities are all listed in Table~\ref{tab:flux}. The first and also the simplest way is to
measure the ratio of the observed fluxes in the LyC to those at a
rest-wavelength of 1500 \AA. This measurement is made after
correcting the fluxes for MW extinction only. The advantage
of this parameter is that it is most directly connected to actual
observational estimation of the rate of escaping ionizing radiation
during the EoR. That is, the observed luminosity density due to star-forming
galaxies at a rest-frame 1500 \AA\  can be measured from the
far-UV luminosity functions during EoR. Knowing the mean ratio of LyC
to 1500 \AA\  fluxes for a representative ensemble of star-forming
galaxies (from observations of lower-z analogs) yields an estimate
of the LyC luminosity density produced by the EoR galaxies. This quantity, $F_{900}/F_{1500}$, for the three
{[SII]}-weak star-forming galaxies are listed in
Table~\ref{tab:flux}. For $F_{1500}$ we fit a simple low-order polynomial 
to the continuum between 1100 and 1500 \AA\ rest-frame and use the
resulting value at 1500 \AA\  since the data are noisy at this wavelength.

Next, we calculate what is sometimes referred to as the relative escape
faction, $\frel$. This is essentially the ratio of the observed flux decrement
across the Lyman break (after correction for MW and internal
extinctions) compared to the intrinsic decrement in the best-fit SB99
model spectrum. As such, the value of the relative escape fraction
is independent of the effects of dust extinction, and is probing only
radiative transfer effects associated with the photo-electric absorption
of the LyC due to hydrogen.

In our specific case we define $\frel$ as:
\begin{equation}
\frel = \left( \frac{F_{910^-}}{F_{910^+}} \right)_{\rm obs} \left( \frac{F_{910^+}}{F_{910^-}}\right)_{\rm int}
\end{equation}
where $F_{910^-}$ is the average extinction-corrected flux densities taken between
rest-frame $\sim$ 885 and 910 \AA\ (again, the exact spectral windows are listed in Table~\ref{tab:flux}),
and $F_{910^+}$ is the average taken between 1050 and 1150 \AA. The latter choice is made to avoid the effects of the $\lya$ airglow
line and the confluence of the high-order Lyman series lines near the Lyman edge.

Finally, we note that dust can be a significant source of opacity
for both ionizing and non-ionizing far-UV radiation in galaxies. We therefore
measure what is commonly referred to as the absolute escape fraction, $\fabs$
(the ratio of emergent LyC flux to the intrinsic flux, including the
effect of dust extinction). Conventionally this is calculated as
\begin{subequations}
\label{subeq:fabs}
\begin{equation}
\fabs=\frel10^{-0.4A_{910}}
\end{equation}
where 
\begin{equation}
A_{910}=\kappa(910\angstrom)\ebvint
\end{equation}
\end{subequations}
is the absorption at 910 \AA.
We obtain $\kappa(910\angstrom)$ by extrapolating the fitting formulae provided in \cite{Calzetti2000,Reddy2015,Reddy2016} slightly towards short wavelength, since the original formulae end at 1200 and 915 \AA, respectively. 

A major source of systematic uncertainty in Equation~\ref{subeq:fabs} is in the UV extinction. To assess this we compare the values for the escape fractions based on the extinction laws adopted by \cite{Calzetti2000} and \cite{Reddy2015,Reddy2016} (see Table~\ref{tab:fesc}). 
There we see that the effects are modest but noticeable; hence we adopt a second approach to circumvent this uncertainty. We use $\sfrir$ to predict the LyC flux in the best-fit SB99 model, and then compare this to the observed LyC flux corrected only for the MW extinction. This quantity is listed in the last column in Table~\ref{tab:fesc}.

In addition, there are systematic uncertainties in escape fraction
associated with the intrinsic Lyman break in the SB99 models. Therefore we compare
the values for both solar and 1/7 solar metallicity models, for burst
ages of $10^{7}$ and $10^{8}$ years, and for models with and without
stellar rotation employed (i.e. \verb|Geneva v40| and \verb|Geneva v00|,
respectively). For completeness, we list the Lyman-break amplitudes defined as the ratio between the
average flux density over 1050-1150 \AA\  and that over 900-910 \AA\  
for different SB99 models in Table~\ref{tab:sb99_models}. The largest
variation is with burst duration. The values we quote for the relative and absolute escape fractions for J0910 and J1432
are obtained from SB99 models with a constant SFR for $10^{7}$ years, while for J1242, they are from SB99 models with a constant SFR for $10^{8}$ years. Those spectra better fit the OVI and NV wind lines. Taking an older burst age for the former two would increase
the escape fractions by $\sim$ 0.2 dex (pushing the relative escape fractions above 1).

\begin{deluxetable*}{ccccc}
\tablecaption{Lyman-break amplitudes, $F_{(1050-1150\mathring{A})}/F_{(900-910\mathring{A})}$, for different SB99 models.
The values used in calculating $f_{\rm esc}$ are indicated with asterisks.
\label{tab:sb99_models}}
\tablecolumns{5}
\tablewidth{0pt}
\tablehead{
\colhead{Name} &
\colhead{$\zsolar$, no rotation} &
\colhead{$\zsolar$, rotation} &
\colhead{$\zsubsolar$, no rotation} &
\colhead{$\zsubsolar$, rotation}
}
\startdata
J0910 ($10^7$ yr) & 2.084 & \, $1.736^\ast$ & 1.792 & 1.756 \\
J0910 ($10^8$ yr) & 3.268 & 2.925 & 2.682 & 2.830 \\
\hline
J1432 ($10^7$ yr) & 2.083 & \, $1.735^\ast$ & 1.792 & 1.756 \\
J1432 ($10^8$ yr) & 3.268 & 2.923 & 2.684 & 2.830 \\
\hline
J1242 ($10^7$ yr) & 2.084 & 1.736 & 1.792 & 1.756 \\
J1242 ($10^8$ yr) & 3.268 & \, $2.924^\ast$ & 2.683 & 2.829 \\
\enddata
\end{deluxetable*}

\section{Discussion\label{sec:discussion}}

In this section we will place the leaky {[SII]}-weak galaxies in
context. First, we will compare their properties to those of the leaky
Green Pea galaxies, which comprise a large majority of the confirmed
low-z leaky galaxies. We will then compare the properties of all the
known low-z leaky galaxies to non-leaky low-z starburst galaxies.
This will allow us to assess the robustness of the various proposed
indirect signposts of leaky galaxies. Finally, we will compare the
properties of the {[SII]}-weak leaky galaxies to leaky galaxies
at $z\sim$ 3 to 4.

\subsection{Comparisons of {[SII]}-Weak and Green Pea Leaky Galaxies}

\begin{figure*}
\gridline{
\fig{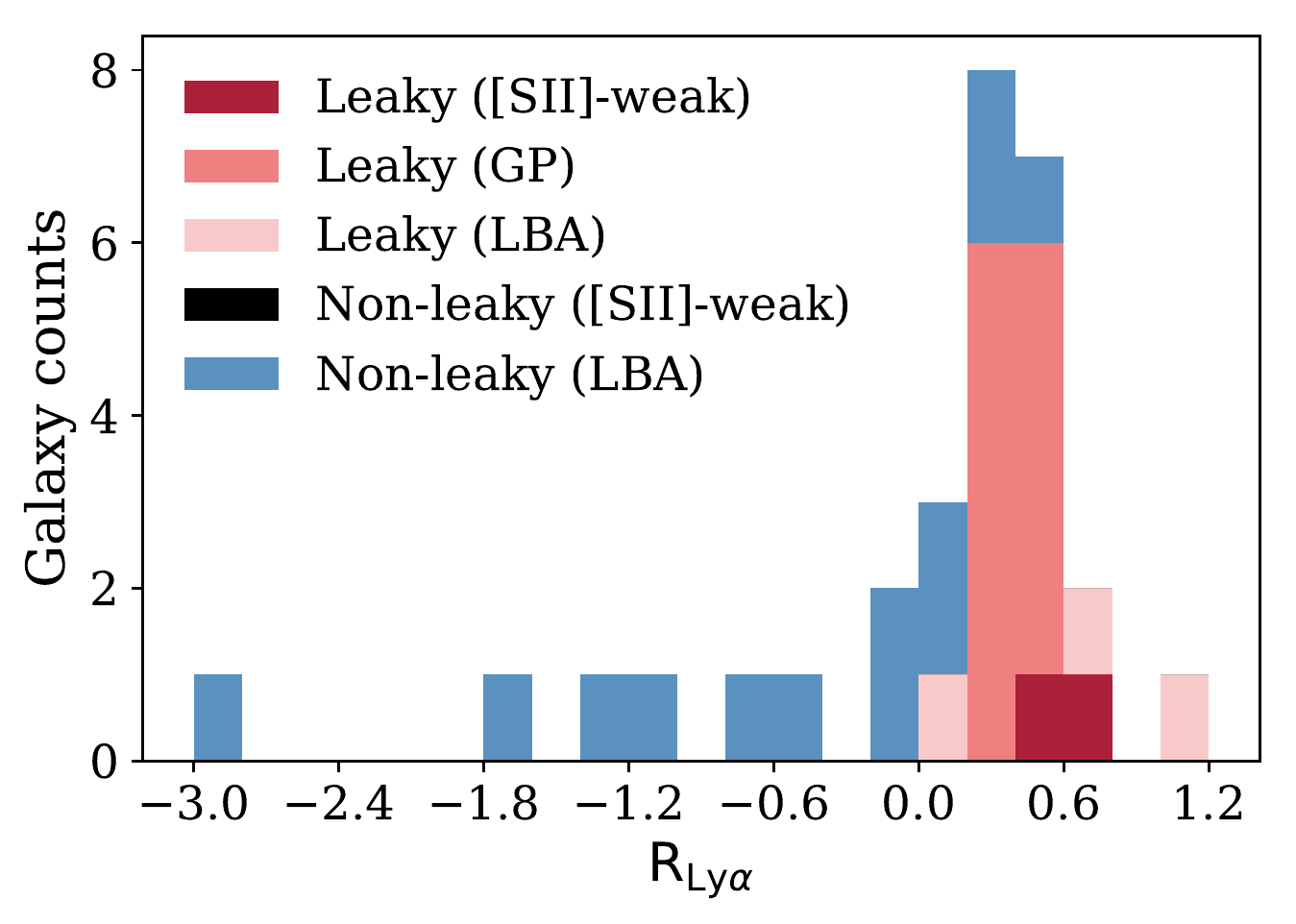}{0.33\textwidth}{(a)}
\fig{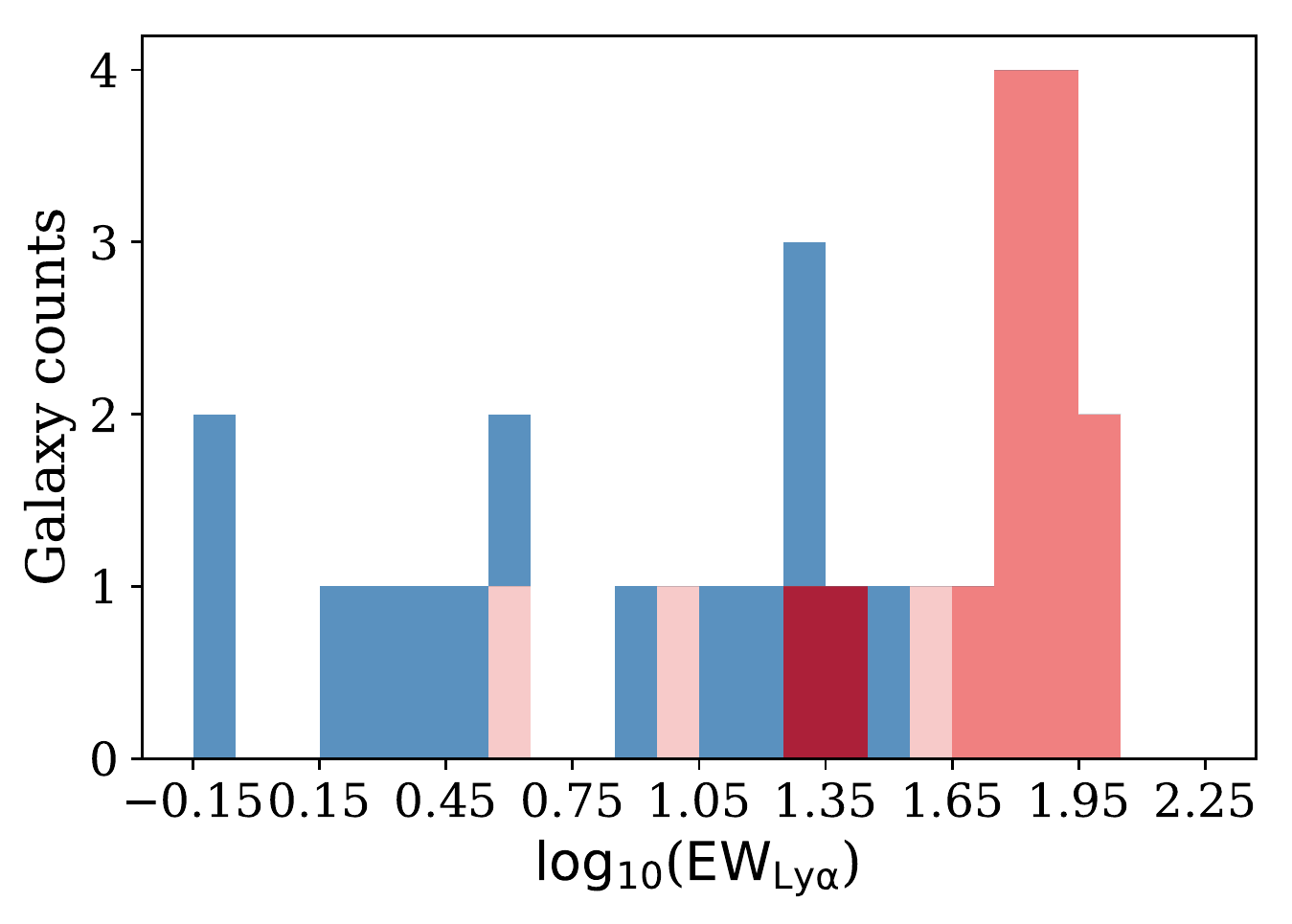}{0.33\textwidth}{(b)}
\fig{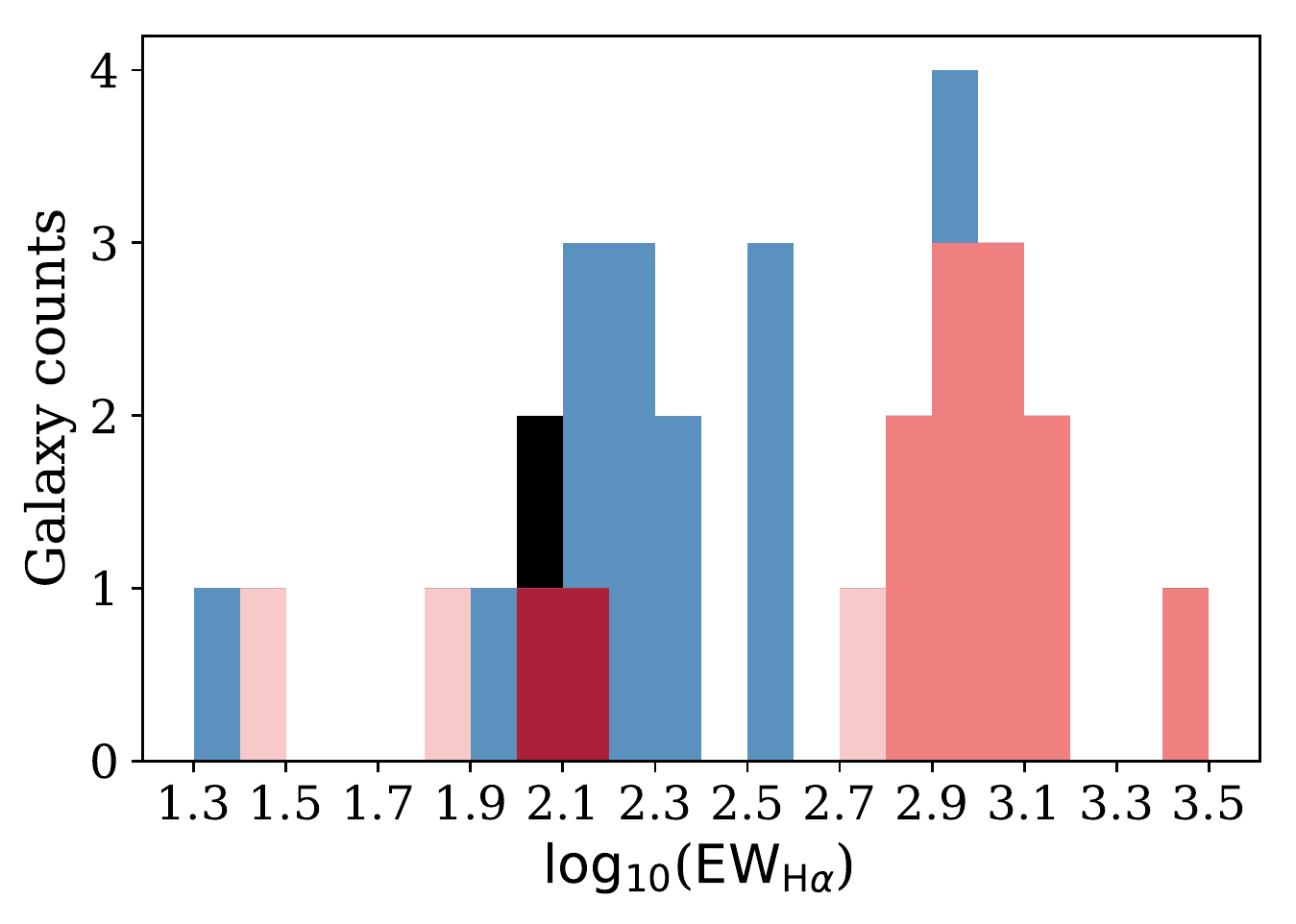}{0.33\textwidth}{(c)}
}
\gridline{\fig{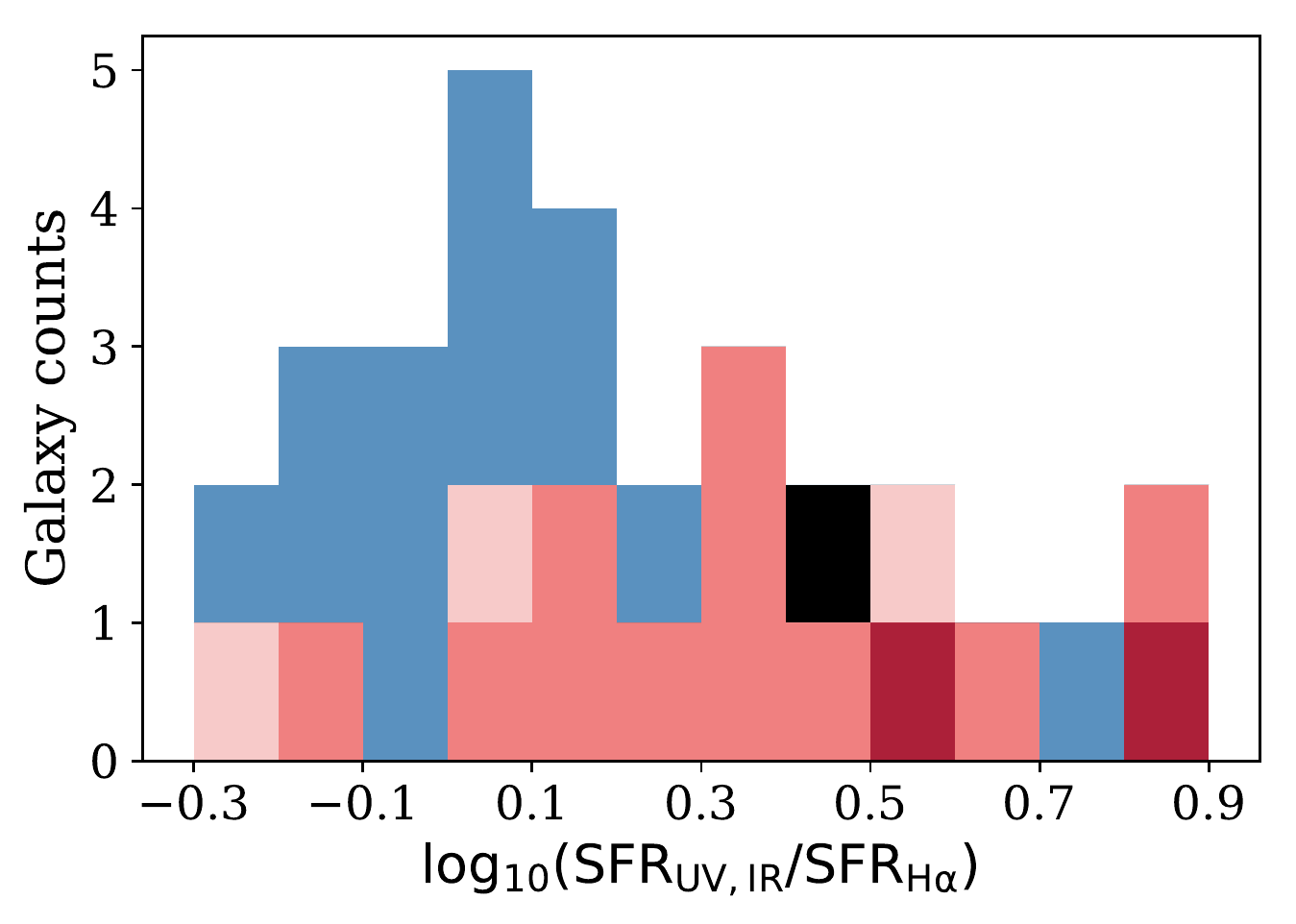}{0.33\textwidth}{(d)}
 \fig{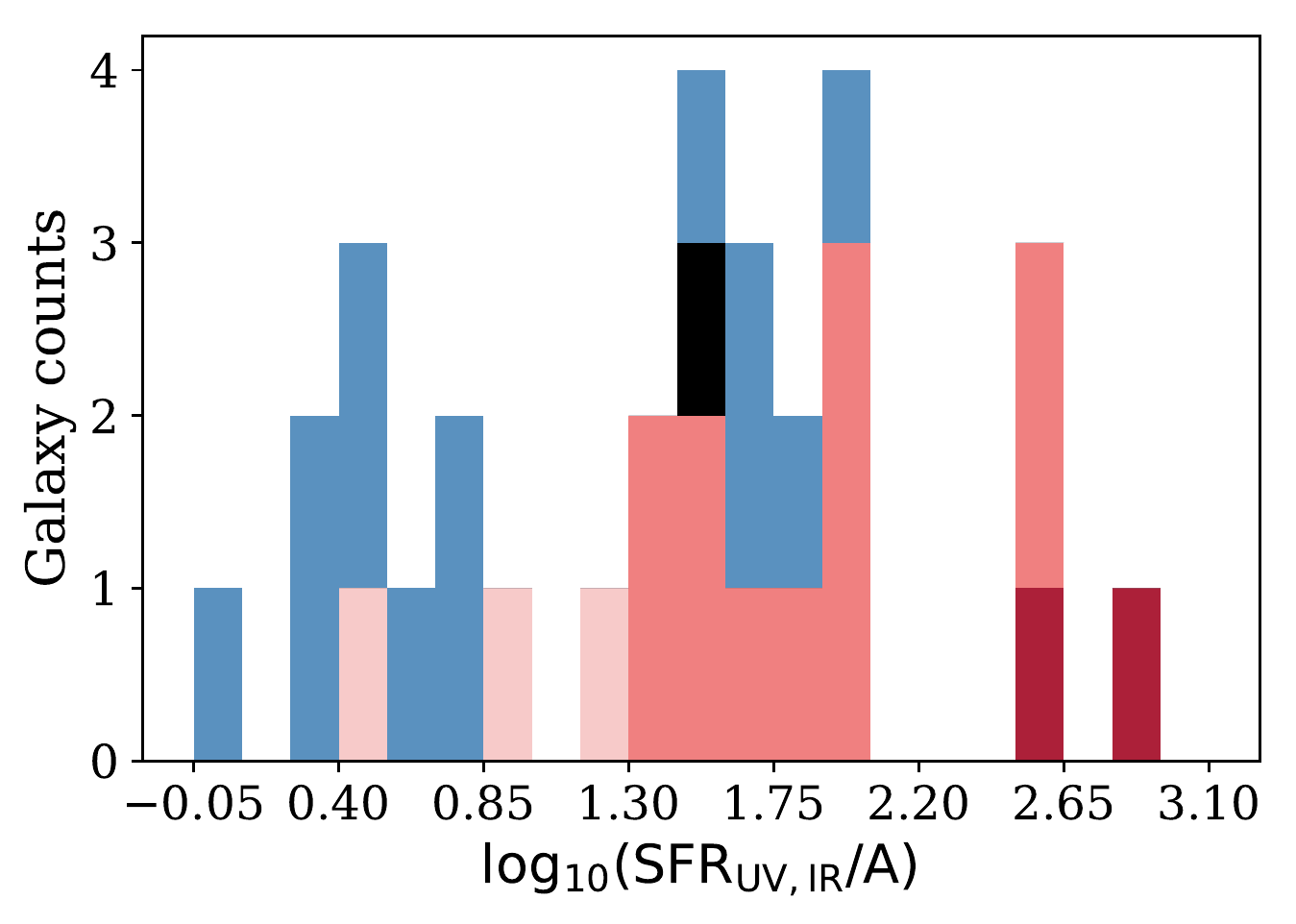}{0.33\textwidth}{(e)}
\fig{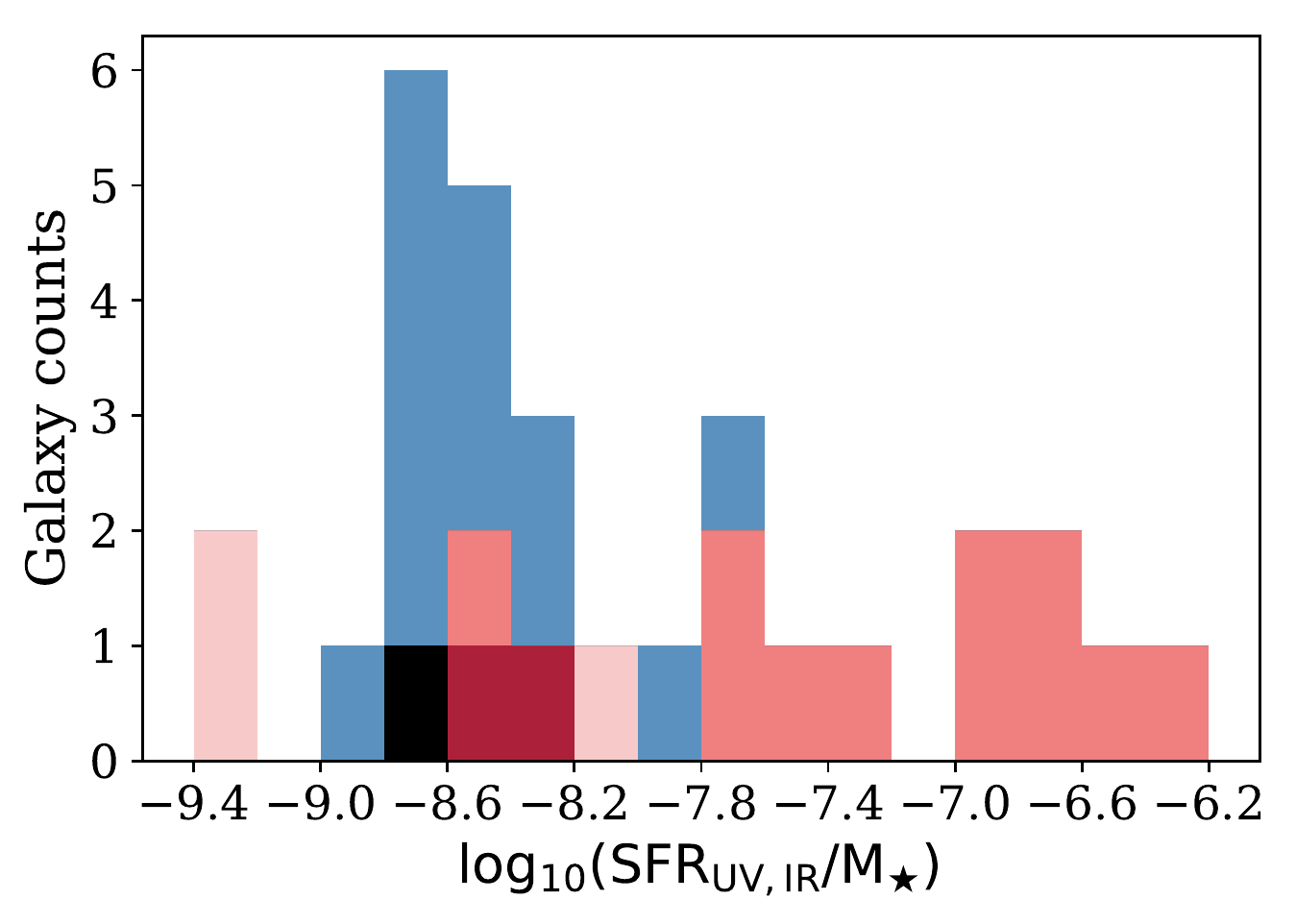}{0.33\textwidth}{(f)}
}
\gridline{
\fig{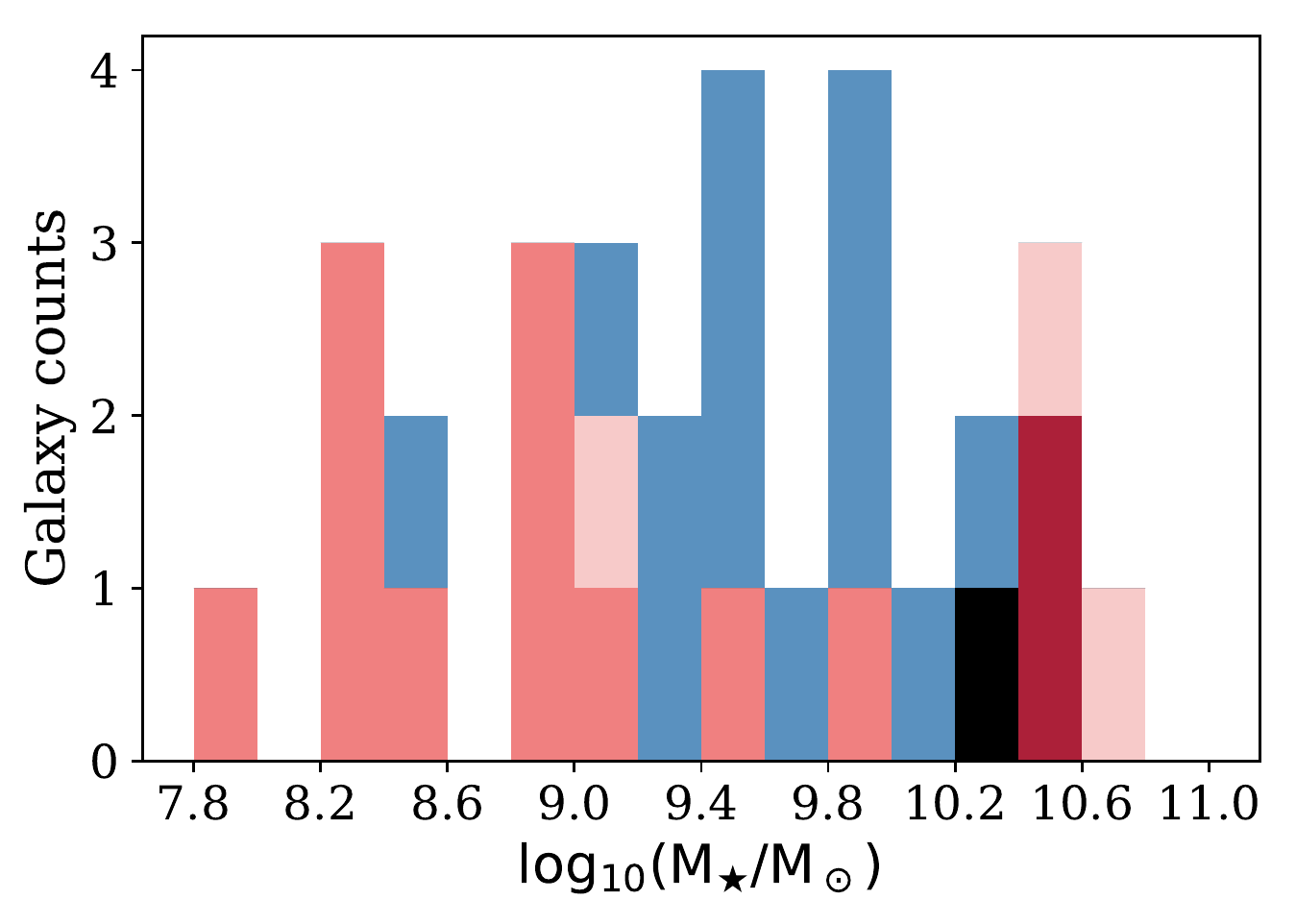}{0.33\textwidth}{(g)}
\fig{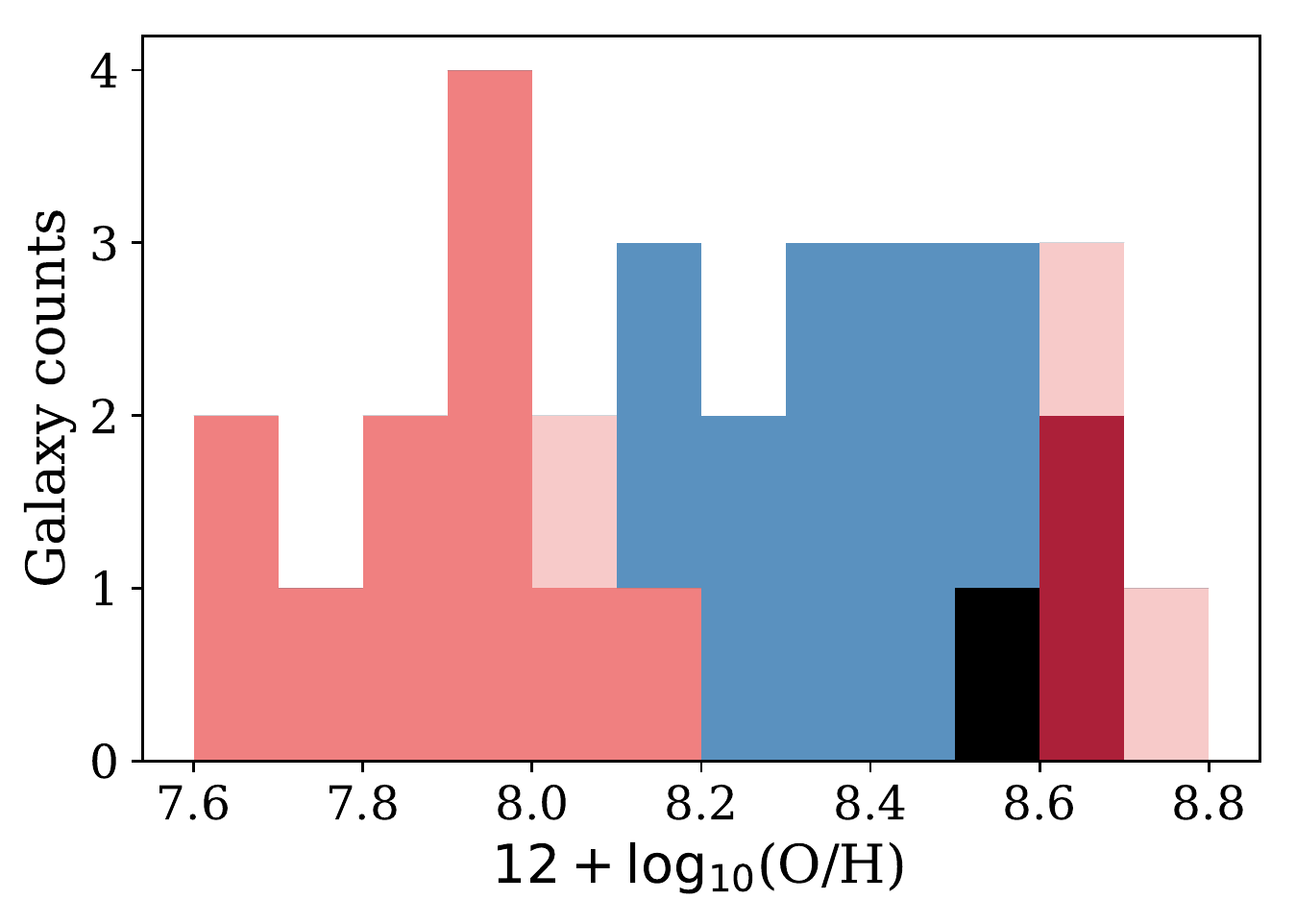}{0.33\textwidth}{(h)}
}
\caption{Histograms of various characteristics of the low-z galaxy samples considered in this paper. Measurements of the three {[SII]}-weak galaxies are tabulated in Table~\ref{tab:measured}. We also provide those of the Green Peas and of the Lyman Break Analogs in Tables~\ref{tab:izotov} and~\ref{tab:ra} respectively in the appendix.\label{fig:hists}}
\end{figure*}

For the Green Pea galaxies, $\mstar$, {[OIII]}/{[OII]}, 
$\ewha$ are taken from the respective references, while the remaining
quantities are calculated the same way as presented in Section~\ref{subsec:Measured-Ancillary-Parameters}
for consistency. Specifically, for $\sfrha$ we estimate the luminosity
of $\ha$ to be used in Equation~\ref{eq:sfr_ha} as 2.86$L_{{\hb}}$,
where $L_{{\hb}}$ is taken from the references; for the $\sfruv$,
we retrieve their COS spectra from MAST, and deredden them using the
reddening law of \cite{Calzetti2000}. Since the Green Peas are nearly dust-free, this calculation of $\sfruv$ is subject to less systematic uncertainty due to internal extinction correction.
These properties are listed in Table~\ref{tab:izotov} in the appendix, and the corresponding
histograms are shown in Figure~\ref{fig:hists}.

As seen in Figure~\ref{fig:hists}, one major difference between the {[SII]}-weak
and Green Pea samples is the stellar mass: the median masses are $10^{8.8}$
and $10^{10.4} \, \msun$ for the Green Peas and {[SII]}-weak galaxies
respectively. This large difference in mass leads to a difference
in gas-phase metallicity: median values of $\rm 12 + log_{10}(O/H)$ of 8.6 and
7.9 for the {[SII]}-weak and Green Peas samples, where a value of
8.7 corresponds to solar metallicity.

The {[SII]}-weak galaxies have extraordinarily high values of SFR/area
(mean of 550 $\msun{\rm yr}^{-1}{\rm kpc}^{-2}$), compared
to a median value of about 75 $\msun{\rm yr}^{-1}{\rm kpc}^{-2}$
for the Green Peas. In terms of SFR/$\mstar$, the Green Peas are
more extreme (median value $10^{-7}\,{\rm yr}^{-1}$, about an order-of-magnitude
larger than the values for the {[SII]}-weak galaxies. This is consistent
with the significantly lower values of the $\ha$ equivalent widths
in the {[SII]}-weak galaxies compared to the Green Peas, and together
these two results suggest that the current bursts in the {[SII]}-weak
galaxies are occurring in the presence of more significant prior star-formation
on timescales longer than a few Myr compared to the Green Peas.

Other emission-line properties of the {[SII]}-weak galaxies are
also much less extreme that those of the Green Peas. As with $\ha$,
the $\lya$ equivalent widths of the {[SII]}-weak galaxies are smaller
than those of the Green Peas by a factor of $\sim$ 3 (23 \emph{vs.} 75 \AA). 
Moreover, as seen in Figure~\ref{fig:o3o2_s2}, the {[SII]}-weak 
galaxies do not exhibit the extraordinarily high ionization level
that is characteristic of the Green Peas (with median {[OIII]}/{[OII]}
fluxes ratios of 1.4 \emph{vs.} 8.0 respectively).

In summary, the {[SII]}-weak galaxies differ significantly in many
of their key properties from the Green Peas: they are 
more massive and more metal-rich, are less-dominated by stars formed in the last few Myr, have a considerably lower ionization state, and have lower absolute LyC escape fractions.

\subsection{Signposts of Leakiness}

There are a number of galaxy characteristics that have been previously
identified as potential signposts of LyC-leakage from galaxies. In
this section we evaluate these signposts in light of our discovery
of this new class of leaky galaxy. To do so, we assemble a sample
of known leaky galaxies at low-z and compare their properties to a
control sample of strong starbursts at similar redshifts that are
unlikely to be leaky. For the sake of consistency, we include only
galaxies with COS data and with the set of galaxy parameters that
can be measured using the spectra in the SDSS.

These samples are drawn from the union of the [SII]-weak galaxies presented in
this paper, the leaky Green Peas in \cite{Izotov2016,Izotov2016b,Izotov2018,Izotov2018a},
and the Lyman Break Analogs in \cite{Alexandroff2015}.
In the latter sample, J0921 has been directly detected below the
Lyman edge \citep{Borthakur2014}. For the other sample members, we
use the residual intensity in the $\lyb$ absorption-line as an indicator
of leakiness, following the results in \cite{Chisholm2018}, and
see also \cite{Steidel2018}. This adds J0213 and J0926 as leakers,
with the 13 other galaxies in Table~\ref{tab:ra} being classified as non-leaky.
\cite{Alexandroff2015} list all the relevant quantities, except
for {[OIII]}/{[OII]}, which we calculate using fluxes obtained
from the MPA-JHU catalog. We note that our definition for {[SII]}-deficit
differs from that in \cite{Alexandroff2015} by taking the horizontal
displacement from the parametric ridge-line as shown in Figure~\ref{fig:sii}
instead of the perpendicular distance between each galaxy and the
ridge-line, so measurements of $\Delta$[SII] are also re-made according to our definition.

We have already compared some of the proposed signposts in the {[SII]}-weak
and Green Pea galaxies in the previous section. In Figure~\ref{fig:hists},
we see that the class of leaky galaxies as-a-whole has somewhat larger
values for SFR/area than the non-leaky starbursts (median values of
51 \emph{vs.} 6 $\msun{\rm yr}^{-1}{\rm kpc}^{-2}$). The leaky
galaxies are more extraordinary in this regard when compared to typical
low-z star-forming galaxies, which have an SFR/Area of only $\sim10^{-2} \, \msun{\rm yr}^{-1}{\rm kpc}^{-2}$
\citep{Kennicutt2012}. We also see that the leaky galaxy sample has
a significantly higher median value for the $\lya$ equivalent width
than the non-leaky galaxies (65 and 4 \AA\ respectively).

Another common property of the leaky galaxies
is that they have a significant amount of blue-shifted $\lya$ emission
(with median value for the $\rlya$ parameter of 0.4 for leaky sample
\emph{vs.} 0.0 for the non-leaky sample. \added{Recently the $\lya$ profiles and their implication for the escape of LyC in Green Peas is discussed in \cite{2019arXiv190809763J}.}
There is also a trend for the leaky galaxies to have significantly higher SFRs based on the IR luminosity or
the extinction-corrected far-UV luminosity than those based on the extinction-corrected
$\ha$ emission-line luminosity, and larger 
than the values in the non-leaky galaxies (median ratios of 2.3 \emph{vs.} 1.1;
also see Figure 10 in \cite{Overzier2009}).

\begin{figure}
  \centering
    \includegraphics[width=0.47\textwidth]{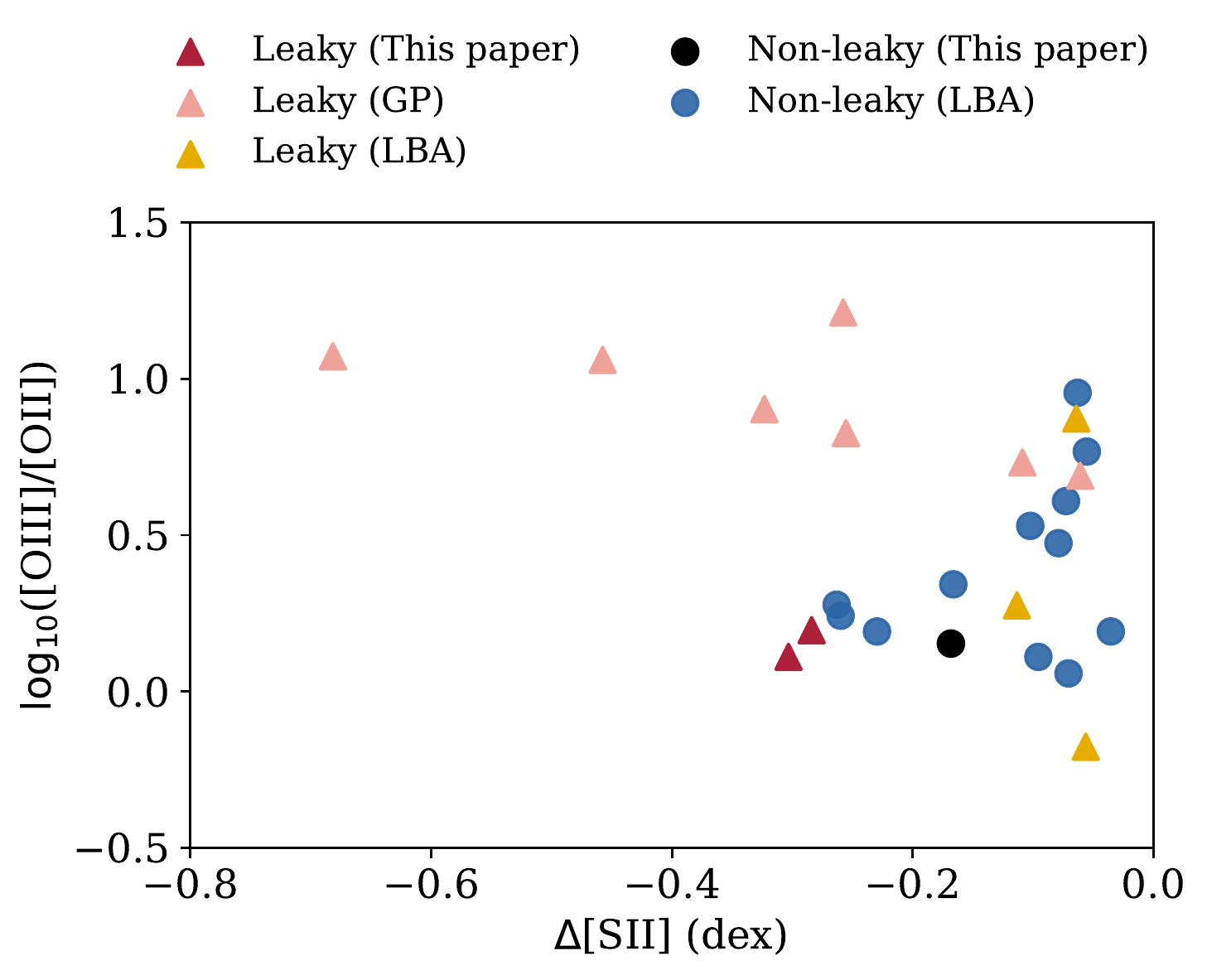}
\caption{Flux ratio of [OIII]5007/[OII]3727 {\it vs.} {[SII]} deficiency for the union of galaxy sample considered in this paper. The two leaky [SII]-weak galaxies are shown as red triangles, while the other non-leaky [SII]-weak galaxy is shown as a black dot. We see that the [SII]-weak leakers do not exhibit the extraordinarily high ionization level that is characteristic of the Green Peas (pink triangles). The remaining galaxies are drawn from Lyman Break Analogs.\label{fig:o3o2_s2}}
\end{figure}

Interestingly, although the leaky Green Peas exhibit a range of $\Delta$[SII] and were not selected based on {[SII]}-weakness, they all have $\Delta$[SII]\textless0. In fact, the five galaxies with the largest {[SII]}-deficiency observed so far in the Lyman continuum (three Green Peas and our two targets) are all leaky (see Figure~\ref{fig:o3o2_s2}).

Thus far, we opt not to discuss in depth any statistical significance which may be manifested in Figure~\ref{fig:hists} due to the still small sample of confirmed leaky galaxies. Rather, we think it is more suitable at present time to describe qualitative trends among the signposts for leakiness to guide future studies.
In light of this, we conclude that the following signposts appear to be robust
(i.e. properties that are in common among the different classes of
low-z leaky galaxies): a high SFR/area, lower values for the SFR measured
from $\ha$ luminosity than from the far-UV plus IR continuum luminosity, strong
$\lya$ emission with a significant fraction that is blue-shifted,
and abnormally weak {[SII]} emission.

All these signposts have physically plausible connections to the escape
of LyC radiation. We have already discussed why {[SII]}-weakness
could be connected to LyC leakage. A high SFR/area leads directly
to a high intensity (flux/area) of ionizing radiation, which can lead
to an ISM that is optically thin to the LyC. It also leads to large
values for radiation pressure and the ram-pressure of a starburst-driven
wind (e.g. \cite{Heckman2015}). The outward forces these generate
can act to expel the ISM and create channels for the escape of ionizing
radiation. As ionizing radiation escapes the ISM, the rate of $\ha$
emission produced by recombination will decrease. A large $\lya$ equivalent
width implies clear channels through which photons resonantly scattered
off HI atoms can escape, and the blue-shifted emission suggests that
we are seeing $\lya$ photons scattered off the near side of an outflowing wind (e.g. \cite{Borthakur2014}).

Finally, it is important to emphasize that these signposts are based on global/isotropic galaxy properties (i.e. properties that should depend only weakly on the observer's particular line-of-sight to the galaxy). This would imply that leakage occurs in a rather isotropic way, instead of just along certain lines-of-sight.

\subsection{The Role of Dominant Central Objects}

We have discussed the evidence above for a general connection between a high SFR/area and the escape of LyC radiation. Here we return to the suggestion in \cite{Heckman2011} and \cite{Borthakur2014} that this escape is made possible by the extreme feedback effects produced by a ``dominant central object'' (DCO). These DCOs were discovered to be present in 20\% of a sample of Lyman Break Analog low-z galaxies imaged with HST \citep{Overzier2009}. They are defined to be compact (marginally resolved by HST), very massive, young objects located at or near the galactic nucleus, and much brighter in the UV than any other star-forming cluster or clump in the galaxy. \cite{Heckman2011} noted that three of the four candidate leaky galaxies in the sample which they analyzed contained a DCO.

As seen in Figure~\ref{fig:acq}, DCOs are present in both of the two leaky {[SII]}-weak galaxies, and produce nearly all the UV emission. In the third (non-leaky) galaxy there is a significant fraction of diffuse UV emission. 
While we do not have robust estimates of the masses of just the DCOs themselves, we can obtain rough values based on the SB99 models for the far-UV spectra since DCOs dominate the far-UV light. The estimated SFRs of 125 and 134 $\msun$ yr$^{-1}$, and ages of $10^7$ years imply that ${\rm M_{\star, DCO}} > 10^9 \, \msun$).
These masses are similar to the values derived from multi-band SED fits to the six DCOs in \cite{Overzier2009}. The measured radii are $\sim$ 300 pc {\it vs.} a mean value of 150 pc for the DCOs in the aforementioned reference. \cite{Overzier2009} showed that the properties of the DCOs are consistent with them being the progenitors of central ``extra light" component found in the centers of cuspy elliptical galaxies, which would have formed during a strong starburst in a dissipative galaxy merger.

\subsection{Comparisons at Higher-Redshift}

\begin{figure}
  \centering
    \includegraphics[width=0.47\textwidth]{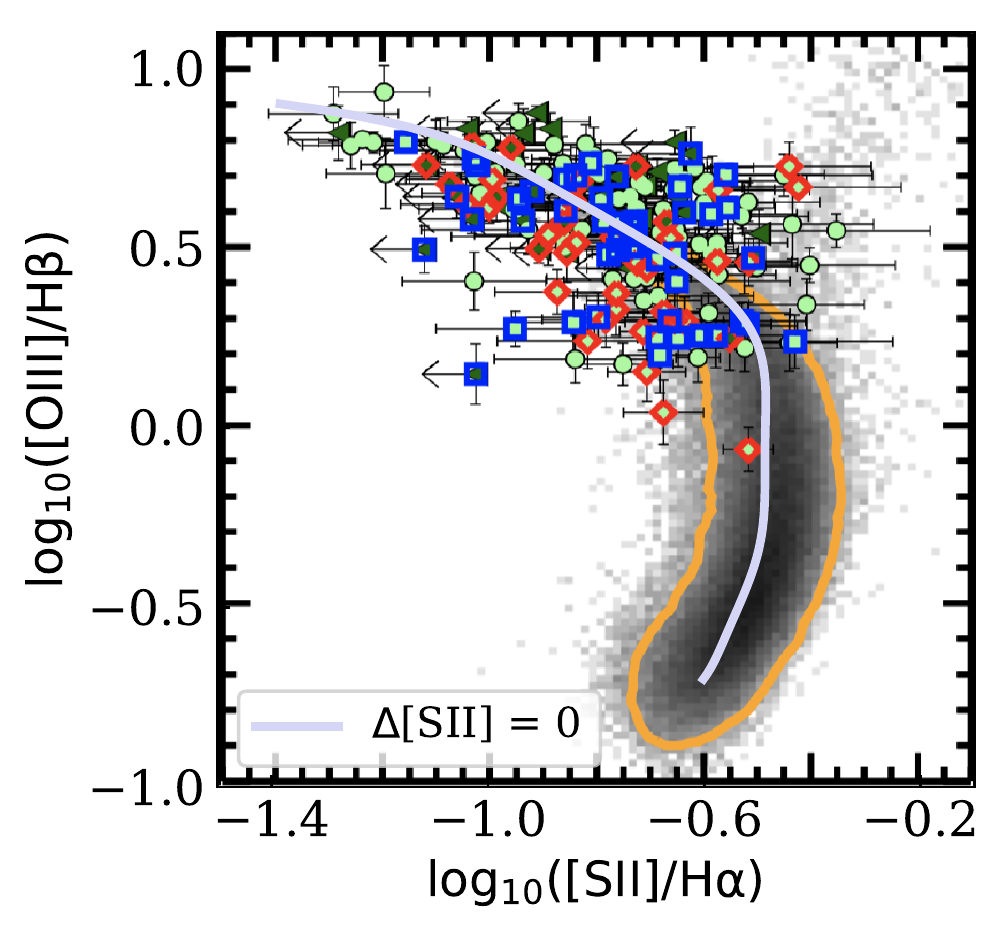}
 \caption{Adapted from Figure~6 of \cite{2018ApJ...868..117S}. The light purple line is our reference line from which {[SII]}-deficiency is quantified. The locus of $z\sim0$ galaxies from SDSS is shown in greyscale, with an orange contour enclosing 90\% of the sample. $\langle z \rangle = 2.3$ galaxies from KBSS are shown as green dots, and galaxies with 2$\sigma$ upper limits on [SII] are shown as dark green triangles.\label{fig:bpt_highz}}
\end{figure}

Before proceeding to further comparisons, we would like to address the validity of our selection criterion when it is extended to higher redshifts. \cite{2018ApJ...868..117S} reported spectral measurements from the Keck Baryonic Structure Survey (KBSS) of about 150 star-forming galaxies at $z\sim$ 2 to 3. Those galaxies fill the upper left region in Figure~\ref{fig:sii} which is sparsely sampled in SDSS (low [SII]/$\ha$ but high [OIII]/$\hb$). We find that the ridge-line in our Figure~\ref{fig:sii} passes right through the center of the data points in Figure~6 of \cite{2018ApJ...868..117S}: see Figure~\ref{fig:bpt_highz}. This shows that the method presented in this paper can be straightforwardly applied to higher redshifts, even though we drew our reference for defining the [SII]-deficiency based on SDSS. It also shows that a minority population of [SII]-weak galaxies are present at these higher-redshifts.

\begin{deluxetable*}{ccccccccccc}
\tablecaption{Comparisons between the mean values calculated from measurements of our two leaky {[SII]}-weak galaxies, and the median of the S18 sample. For the {[SII]}-weak sample, we use the values based on the extinction law in \cite{Reddy2015,Reddy2016}. Unless otherwise noted, the values for the S18 sample are taken directly from S18. The SFR for S18 is based on the bolometric luminosity in S18 and the prescription in \cite{Kennicutt2012}. The value for $\mstar$ assumes that these galaxies lie along the star-forming main sequence \citep{Reddy2012}. The value for $\rlya$ is our estimate based on the published stacked spectrum in S18.\label{tab:s18}}
\tablecolumns{11}
\tablewidth{0pt}
\tablehead{
\colhead{} &
\colhead{$\rm M_{FUV}$} &
\colhead{$\rm log_{10} \mstar$} &
\colhead{$\ebvint$} &
\colhead{$\ewlya$} &
\colhead{$\rlya$} &
\colhead{SFR} &
\colhead{SFR/$\mstar$} &
\colhead{$F_{900}/F_{1500}$} &
\colhead{$\frel$} &
\colhead{$\fabs$} 
\\
\colhead{} &
\colhead{(AB mag)} &
\colhead{($\msun$)} &
\colhead{(mag)} &
\colhead{(\AA)} &
\colhead{} &
\colhead{($\msun {\rm yr^{-1}}$)} &
\colhead{(${\rm Gyr}^{-1}$)} &
\colhead{} &
\colhead{} &
\colhead{} 
}
\startdata
[SII] & -21.4 & 10.5 & 0.25   & 23 & 0.60 & 130 & 4.2 & 0.35 & 0.77 & 0.04 \\
S18 & -20.9 & 9.8   & 0.045 & 28 & 0.35 & 25   & 4.0 & 0.36 & 1.21 & 0.70
\enddata
\end{deluxetable*}

We now compare the properties of the [SII]-weak leaky galaxies to other leaky galaxies at higher redshifts. 
\cite{Steidel2018} (hereafter S18) reported the detection of LyC flux in 15 individual
galaxies at $z\sim3$ (out of sample of 124 galaxies), and in stacked
spectra binned according to various galaxy properties. \cite{Marchi2017}
have observed 401 galaxies at $z\sim4$, and detected LyC flux in
stacks of spectra binned in various ways. \cite{2018MNRAS.476L..15V} reported the highest redshift individually-confirmed LyC-leaky galaxy at $z=4$, and \cite{2019arXiv190407941V} found evidence of a compact region emitting LyC radiation at $z\sim3$.

The results on the properties of these leaky galaxies are qualitatively consistent with the results presented above for the low-z leaky galaxies: a higher escape fraction is associated with compact sizes (radii $<300$ pc) and with strong $\lya$ emission. Since S18 tabulate the median properties of their individual detections, we directly compare these values to those of our two leaky {[SII]}-weak galaxies. This is presented in Table \ref{tab:s18}. 

In many respects, the galaxies in the two samples are similar, including
the properties of the $\lya$ emission-line, the specific SFR, $F_{900}/F_{1500}$,
and $\frel$. The {[SII]}-weak galaxies are somewhat more massive, 
and have higher SFRs. The biggest difference is
in the larger amount of dust extinction in the {[SII]}-weak galaxies,
which leads to smaller absolute escape fractions.
This may reflect higher ($\sim$ solar) ISM metal abundances (higher
dust-to-gas ratio) in the {[SII]}-weak galaxies.

\section{Conclusions\label{sec:conclusions}}

We have reported on observations with COS on HST of three low-z
($z\sim0.3$) starburst galaxies, selected on the basis of the relative
weakness of the {[SII]}6717,6731 nebular emission-lines 
defined with respect to normal star-forming galaxies.
This is a proposed signpost for galaxies that are optically-thin to ionizing radiation.
We detect a significant flux below the Lyman limit in two of the three
galaxies, with relative escape fractions of 93\% and 80\% respectively
and absolute escape fractions of 3\% and 4\%.

We have compared these {[SII]}-weak galaxies to other known classes
of ``leaky'' galaxies. Compared to the low-z Green Peas, the
{[SII]}-weak leaky galaxies have significantly larger stellar masses,
higher metallicities, larger amounts of dust extinction, a much lower ionization state (as traced by the nebular emission-lines), smaller $\lya$ emission-line equivalent widths, and have optical spectra that are less dominated by a very young (few Myr-year old) starburst. 

We have compared the properties
of the entire known set of low-z leaky galaxies to non-leaky starbursts
at similar redshifts. We find that the leaky galaxies have higher
SFR per unit area, stronger $\lya$ emission-lines, and a greater
fraction of the $\lya$ emission produced by blue-shifted material.
Interestingly, we find that while the Green Peas were not selected
based on {[SII]} properties, they too have relatively weak {[SII]}
emission-lines. We also find that leaky galaxies have significantly
lower SFRs based on Balmer emission-line luminosity 
than those based on the intrinsic far-UV plus IR continuum luminosity (as required if a large fraction of ionizing photons escape).

We have also compared the {[SII]}-weak galaxies to samples of leaky
galaxies at $z\sim3$ to 4. We find overall similarities, including
compact sizes and relatively strong $\lya$ emission. Compared to
the sample of galaxies at $z\sim3$ that are individually-detected
in the LyC, the {[SII]}-weak galaxies differ most strongly in having
larger amounts of dust extinction, which results in significantly
smaller values for the absolute escape fraction (even though the relative
escape fractions are similar). This may reflect a higher ($\sim$
solar) ISM metallicity and a correspondingly higher dust/gas ratio
in the {[SII]}-weak galaxies. We have also shown that our technique for selecting [SII]-weak galaxies can be applied out to redshifts $\sim$ 2 to 3, based on existing spectra.

We thus conclude that {[SII]}-weakness is a highly effective way to identify galaxies that are likely to be leaking a significant amount of LyC radiation. Since the leaky galaxies described in this paper are so different from Green Peas, this technique potentially expands the range of galaxy properties over which such searches for leaky galaxies can be done. This will improve our opportunities to use low-z leaky galaxies as local laboratories in which the physical processes and characteristics that allow LyC photons to escape can be investigated. It also suggests that there may be a variety of different physical conditions and processes that make galaxies leaky. Finally, it gives us an additional technique to identify leaky galaxies during the EoR using spectroscopic observations with JWST.

\acknowledgments
B.W. thanks Sihao Cheng and Hsiang-Chih Hwang for valuable conversations, Kate Rowlands for assistances on SDSS data sets, and Weichen Wang for clarifications on dust extinction.
This work is supported by HST-GO-15341, provided by NASA through a grant from the Space Telescope Science Institute, which is operated by the Association of Universities for Research in Astronomy, Inc., under NASA contract NAS5-26555. R.A.O. is grateful for financial support from FAPESP grant 2018/02444-7.
This publication made use of data products from the Wide-field Infrared Survey Explorer, which is a joint project of the University of California, Los Angeles, and the Jet Propulsion Laboratory/California Institute of Technology, funded by the National Aeronautics and Space Administration; the NASA/IPAC Extragalactic Database, which is operated by the Jet Propulsion Laboratory, California Institute of Technology, under contract with the National Aeronautics and Space Administration; and the NASA Astrophysical Data System for bibliographic information.

\vspace{5mm}
\facilities{HST(COS), GALEX, Sloan, WISE}
\software{Astropy \citep{2013A&A...558A..33A, 2018AJ....156..123A}, CalCOS, Matplotlib \citep{Matplotlib}, NumPy \citep{numpy}, SciPy \citep{scipy}, statmorph \citep{statmorph}}

\bibliography{leaky}

\appendix

\begin{deluxetable*}{cccccccccccc}[h]
\tablecaption{Measurements of Green Pea galaxies in~\cite{Izotov2016,Izotov2016b,Izotov2018,Izotov2018a}.\label{tab:izotov}}
\tablecolumns{12}
\tablewidth{0pt}
\tablehead{
\colhead{Name} &
\colhead{$\mstar$} &
\colhead{$r_{50}$} &
\colhead{$\frac{\sfruv}{\sfrha}$}  & 
\colhead{$\sfruv/A$} &
\colhead{$\sfruv/\mstar$} &
\colhead{$\ewlya$} &
\colhead{$\rlya$} &
\colhead{$\ewha$} &
\colhead{${\rm \frac{[OIII]}{[OII]}}$} &
\colhead{$\Delta${[SII]}} &
\colhead{$12+{\rm log_{10} ({\rm \frac{O}{	H}})}$} \\
\colhead{} &
\colhead{(${\rm log_{10}} \msun$)} &
\colhead{(kpc)} &
\colhead{} &
\colhead{($\msun {\rm yr^{-1}}{\rm kpc^{-2}}$)} &
\colhead{(${\rm log_{10}yr^{-1}}$)} &
\colhead{(\AA)} &
\colhead{} &
\colhead{(\AA)} &
\colhead{} &
\colhead{(dex)} &
\colhead{}
}
\startdata
J1152 & 9.59 & 0.49 & 2.33 & 43 & -7.78 & 54.66 & 0.52 & 1320 & 5.4 & -0.11 & 8.0 \\
J1333 & 8.5 & 0.56 & 6.38 & 32 & -6.71 & 60.62 & 0.22 & 817 & 4.8 & - & 7.76 \\
J1442 & 8.96 & 0.25 & 5.01 & 325 & -6.86 & 80.55 & 0.24 & 1122 & 6.7 & -0.26 & 7.93 \\
J1503 & 8.22 & 0.29 & 2.0 & 102 & -6.49 & 69.17 & 0.24 & 1438 & 4.9 & -0.06 & 7.95 \\
\hline
J0925 & 8.91 & 0.35 & 2.32 & 112 & -6.99 & 68.91 & 0.39 & 732 & 5.0 & - & 7.91 \\
\hline
J0901 & 9.8 & 0.37 & 1.57 & 24 & -8.48 & 106.83 & 0.3 & 831 & 8.0 & -0.32 & 8.16 \\
J1011 & 9.0 & 0.13 & 2.63 & 365 & -7.38 & 74.96 & 0.52 & 1052 & 27.1 & - & 7.99 \\
J1243 & 7.8 & 0.24 & 1.99 & 86 & -6.31 & 83.87 & 0.52 & 740 & 13.5 & - & 7.89 \\
J1248 & 8.2 & 0.25 & 1.19 & 75 & -6.72 & 107.54 & 0.47 & 2561 & 11.8 & -0.68 & 7.64 \\
J1256 & 8.8 & 0.24 & 1.39 & 29 & -7.77 & 60.2 & 0.24 & 955 & 16.3 & -0.26 & 7.87 \\
\hline
J1154 & 8.2 & 0.18 & 0.68 & 25 & -7.51 & 86.48 & 0.44 & 1150 & 11.5 & -0.46 & 7.62 \\
\enddata
\end{deluxetable*}

\begin{deluxetable*}{ccccccccccccc}
\tablecaption{Measurements of Lyman Break Analogs in~\cite{Alexandroff2015}. ``$\blacktriangle$" and ``x" stands for ``leaky" and ``non-leaky"
respectively.\label{tab:ra}}
\tablecolumns{13}
\tablewidth{0pt}
\tablehead{
\colhead{Name} &
\colhead{leakiness} &
\colhead{$\mstar$} &
\colhead{$r_{50}$} &
\colhead{$\frac{\sfruv}{\sfrha}$}  & 
\colhead{$\sfruv/A$} &
\colhead{$\sfruv/\mstar$} &
\colhead{$\ewlya$} &
\colhead{$\rlya$} &
\colhead{$\ewha$} &
\colhead{${\rm \frac{[OIII]}{[OII]}}$} &
\colhead{$\Delta${[SII]}} &
\colhead{$12+{\rm log_{10} ({\rm \frac{O}{	H}})}$} \\
\colhead{} &
\colhead{} &
\colhead{(${\rm log_{10}} \msun$)} &
\colhead{(kpc)} &
\colhead{} &
\colhead{($\msun {\rm yr^{-1}}{\rm kpc^{-2}}$)} &
\colhead{(${\rm log_{10}yr^{-1}}$)} &
\colhead{(\AA)} &
\colhead{} &
\colhead{(\AA)} &
\colhead{} &
\colhead{(dex)} &
\colhead{}
}
\startdata
J0055 & x & 9.7 & 0.32 & 0.82 & 36.65 & -8.33 & 2.32 & -1.25 & 375 & 3.38 & -0.1 & 8.28 \\
J0150 & x & 10.3 & 1.37 & 1.88 & 3.17 & -8.73 & 3.04 & -1.72 & 199 & 2.2 & -0.17 & 8.4 \\
J0213 & $\blacktriangle$ & 10.5 & 0.39 & 3.33 & 19.84 & -9.22 & 9.2 & 0.69 & 31 & 1.89 & -0.11 & 8.76 \\
J0921 & $\blacktriangle$ & 10.8 & 0.78 & 1.25 & 7.68 & -9.33 & 4.01 & 1.04 & 72 & 0.67 & -0.06 & 8.69 \\
J0926 & $\blacktriangle$ & 9.1 & 0.69 & 0.59 & 3.47 & -8.08 & 36.22 & 0.14 & 577 & 7.47 & -0.06 & 8.05 \\
J1025 & x & 9.2 & 0.61 & 0.62 & 3.23 & -8.32 & 20.71 & 0.02 & 395 & 5.85 & -0.06 & 8.11 \\
J1112 & x & 10.2 & 0.33 & 1.16 & 41.9 & -8.74 & 7.6 & -0.63 & 205 & 1.75 & -0.26 & 8.52 \\
J1113 & x & 9.6 & 1.09 & 5.67 & 0.95 & -8.75 & 0.85 & -0.09 & 24 & 1.14 & -0.07 & 8.35 \\
J1144 & x & 9.9 & 0.76 & 1.26 & 2.45 & -8.95 & 0.78 & -2.89 & 85 & 1.56 & -0.04 & 8.4 \\
J1414 & x & 8.5 & 0.63 & 0.81 & 2.06 & -7.79 & 1.83 & 0.28 & 351 & - & - & - \\
J1416 & x & 10.0 & 0.19 & 1.17 & 102.94 & -8.63 & 1.69 & 0.5 & 183 & 1.89 & -0.26 & 8.47 \\
J1428 & x & 9.6 & 0.71 & 0.7 & 4.39 & -8.46 & 19.65 & 0.04 & 249 & 2.98 & -0.08 & 8.31 \\
J1429 & x & 9.4 & 0.29 & 0.74 & 50.72 & -7.97 & 32.17 & 0.27 & 850 & 9.01 & -0.06 & 8.12 \\
J1521 & x & 9.5 & 0.37 & 0.98 & 6.8 & -8.73 & 3.96 & -1.07 & 145 & 4.06 & -0.07 & 8.27 \\
J1525 & x & 9.4 & 0.51 & 1.43 & 5.54 & -8.44 & 16.57 & -0.01 & 126 & 1.29 & -0.1 & 8.46 \\
J1612 & x & 10.0 & 0.31 & 1.12 & 59.87 & -8.44 & 13.6 & -0.41 & 174 & 1.55 & -0.23 & 8.51 \\
\enddata
\end{deluxetable*}

\end{document}